\documentclass[12pt]{aastex}
\usepackage{natbib}
\usepackage{threeparttable}
\usepackage{subfigure}
\usepackage{amssymb,amsmath}
\usepackage{hyperref}
\usepackage{graphicx}
\usepackage{epstopdf}

\begin{document}
\pagestyle{plain}
\pagenumbering{arabic}

\title{Measurements of Coronal Faraday Rotation at 4.6 Solar Radii}
\author{Jason E. Kooi, Patrick D. Fischer, Jacob J. Buffo, and Steven R. Spangler}
\affil{Department of Physics and Astronomy, University of Iowa}
\affil{Iowa City, IA, 52240, USA}
\email{jason-kooi@uiowa.edu}


\begin{abstract}
Many competing models for the coronal heating and acceleration mechanisms of the high-speed solar wind depend on the solar magnetic field and plasma structure in the corona within heliocentric distances of $5R_\odot$.  We report on sensitive VLA full-polarization observations made in August, 2011, at 5.0 and 6.1 GHz (each with a bandwidth of 128 MHz) of the radio galaxy 3C228 through the solar corona at heliocentric distances of $4.6-5.0R_\odot$. Observations at 5.0 GHz permit measurements deeper in the corona than previous VLA observations at 1.4 and 1.7 GHz.  These Faraday rotation observations provide unique information on the magnetic field in this region of the corona.  The measured Faraday rotation on this day was lower than our a priori expectations, but we have successfully modeled the measurement in terms of observed properties of the corona on the day of observation.  Our data on 3C228 provide two lines of sight (separated by $46\arcsec$, 33,000 km in the corona).  We detected three periods during which there appeared to be a difference in the Faraday rotation measure between these two closely spaced lines of sight.  These measurements (termed differential Faraday rotation) yield an estimate of $2.6$ to $4.1$ GA for coronal currents.  Our data also allow us to impose upper limits on rotation measure fluctuations caused by coronal waves; the observed upper limits were $3.3$ and $6.4$ rad/m$^2$ along the two lines of sight.  The implications of these results for Joule heating and wave heating are briefly discussed.
\end{abstract}



\section{Introduction}\label{sec:Intro}

Knowledge of the coronal magnetic field, its magnitude and form as a function of heliocentric coordinates, is crucial for understanding a number of phenomena in space plasma physics and astrophysics, such as the heating of the solar corona, the acceleration of the solar wind, and the plasma structure of the interplanetary medium.  Unfortunately, there are limited observational tools for determining its strength as a function of heliocentric distance $r$,  and even fewer for providing further information.  Illustrative references to what is known are \cite{Dulk&McLean:1978}, \cite{Bird&Edenhofer:1990}, \cite{Mancuso:2003}, and \cite{Bird:2007}.  


\subsection{Coronal Heating Mechanisms}\label{sec:Heating_Models}

One of the biggest mysteries in solar physics is the identity of the mechanism that heats the corona and drives the solar wind.  Proposed heating mechanisms can be divided into two groups: direct current (DC) and alternating current (AC) heating.  DC heating is the dissipation of magnetic energy in the conventional Joule heating process; AC heating, or wave heating, is thought to be caused by wave energy dissipation.  An alternative suggestion by \cite{Scudder:1992a,Scudder:1992b} proposed that non-thermal electron and/or ion velocity distributions with suprathermal tails can reproduce the observed temperature profile of the coronal base ($\sim1R_\odot$) as well as the scale length of the transition region and argued that an ad hoc heating mechanism is unnecessary.  Most models that do propose a heating mechanism, though, generally focus on heating at the (1) coronal base ($1-1.5R_\odot$) or (2) extended radial distances at and beyond the point where the fast solar wind becomes supersonic ($2-5R_\odot$).  Faraday rotation measurements have provided the best measurements of magnetic field fluctuations within $3-15R_\odot$ and have been used to place constraints on these models.

DC heating models of the solar corona invoke Joule heating from coronal currents, thought to be contained in turbulent current sheets.  This idea originated with \cite{Parker:1972} and has been elaborated in many subsequent works \cite[see, e.g.,][]{Gudiksen&Nordlund:2005,Peter:2006}.  However, there are few observational measurements of coronal currents.  \cite{Spangler:2007,Spangler:2009} developed a method of measuring the magnitudes of coronal currents using differential Faraday rotation measurements (the difference between Faraday rotation along two or more closely spaced lines of sight).  This method was motivated by \cite{Brower:2002} and \cite{Ding:2003} who used differential Faraday rotation measurements to determine the internal currents in a laboratory plasma, the Madison symmetric torus (MST) reversed-field pinch (RFP) at the University of Wisconsin \citep{Prager:1999}.  \cite{Spangler:2007} found detectable differential Faraday rotation yielding estimates of $2.3\times10^8$ to $2.5\times10^9$ A and concluded that these measured currents were irrelevant for coronal heating unless the true resistivity in the corona exceeds the Spitzer value by several orders of magnitude.

AC heating models presume that the characteristic driving time scales are short compared to the transit time (e.g., for acoustic or Alfv{\' e}n waves) across coronal structures \citep{Cranmer:2002}.  The resulting wavelike oscillations damp out as they enter the corona and provide the necessary heating.  In these models, Alfv{\' e}n waves play the central role.  They are a major component of spatial and temporal variations in the solar wind and have large amplitudes and long periods (hours) in regions well beyond the corona where direct in situ measurements can be made \citep{Hollweg:2010}.  If the power contained within these waves can be efficiently dissipated (e.g., a turbulent cascade to dissipation scales), they could heat the corona and power the solar wind.  Wave dissipation models for coronal heating and solar wind generation require a wave flux density emerging from the coronal base in the range $2-5\times10^5$ erg$-$s$^{-1}-$cm$^{-2}$ \cite[see, e.g.,][]{Hollweg:1986}.  \cite{Hollweg:1982} had previously shown that Faraday rotation fluctuations were associated primarily with fluctuations in the magnetic field and that the wave flux is sufficiently great to drive coronal heating.  \cite{Efimov:1993}, \cite{Andreev:1997a}, and \cite{Hollweg:2010} also concluded that {\it Helios} Faraday rotation fluctuations were consistent with coronal magnetic field fluctuations with sufficient amplitudes to power the solar wind.

\cite{Sakurai&Spangler:1994a} measured a mean magnetic field of $12.5\pm2.3$ mG at $\sim9R_\odot$, in reasonable agreement with extrapolations from the {\it Helios} measurements, but determined the wave flux at the coronal base to be less than $1.6\times10^5$ erg$-$s$^{-1}-$cm$^{-2}$.  \cite{Mancuso&Spangler:1999} reported on similar measurements made of the radio galaxy J0039+0319, showing that the inferred wave flux at the coronal base ranges from $2.4\times10^4$ to $2.3\times10^5$ erg$-$s$^{-1}-$cm$^{-2}$.  Further, they demonstrated that the typical scale of waves causing Faraday rotation fluctuations is at least on the order of $0.15R_\odot$ ($\sim10^5$ km) and may be larger up to an order of magnitude.  In order for the waves to efficiently heat the corona, the wave power must be on the scale of the ion Larmor radius (on the order of kilometers at a heliocentric distance of $5R_\odot$).  These two sets of results thus differ mildly regarding the conclusion as to whether the observed wave flux in the corona is quantitatively adequate to explain the heating of the corona and the acceleration of the solar wind.  \cite{Hollweg:2010} used the {\it Helios} observations instead of the data from natural radio sources to investigate the MHD wave-turbulence model of \cite{Cranmer:2007} because the lines of sight sampled numerous coronal occultations over a broad range of heliocentric distances ($2-15R_\odot$) during solar minimum ($1975-1977$).



\subsection{Faraday Rotation}\label{sec:FR_intro}

This paper will deal with coronal probing via Faraday rotation of radio waves from an extragalactic radio source.  An advantage of Faraday rotation is that it provides a measurement of the polarity as well as magnitude of the magnetic field.  Further, Faraday rotation is one of the few diagnostic tools with which to probe the magnetic field and field fluctuations of the corona.  At the photosphere, the vector magnetic field can be measured by Zeeman splitting of spectral lines.  The vector magnetic field has also been measured in situ by spacecraft, but only outside of $\sim62 R_\odot$ (e.g., {\it Helios}).  Within heliocentric distances of $<30R_\odot$, the Zeeman effect is typically much smaller than thermal broadening effects and therefore very difficult to measure.  However, it is within this region (specifically, beyond $1.5R_\odot$) that the solar wind accelerates: the fast component of the solar wind accelerates from sub- to supersonic speeds before reaching a terminal velocity typically near $\sim10R_\odot$ \citep{Grall:1996} and the slow component is still accelerating as far out as $\sim40R_\odot$ \citep{Bird:1994}.  It is also within this region that some wave-driven models predict heat addition to the corona.

Faraday rotation is a change in the polarization position angle $\chi$ of polarized radiation as it propagates through a magnetized plasma.  The change in position angle, or rotation, $\Delta \chi$ is given by 
\begin{equation}\label{eq:FR}
\Delta \chi = \left[ \left( \frac{e^3}{2 \pi m_e^2 c^4}\right) \int_{LOS} n_e \mathbf{B} \cdot \mathbf{ds} \right] \lambda^2
\end{equation}
in cgs units.  The variables in Equation~\eqref{eq:FR} are as follows.  The fundamental physical constants $e, m_e, \mbox{ and } c$ are, respectively, the fundamental charge, the mass of an electron, and the speed of light.  The term in parentheses has the numerical value $C_{FR} \equiv 2.631\times10^{-17}$ rad$-$G$^{-1}$.  The electron density in the plasma is $n_e$ and $\mathbf{B}$ is the vector magnetic field.  The incremental vector $\mathbf{ds}$ is a spatial increment along the line of sight, which is the path on which the radio waves propagate. Positive $s$ is in the direction from the source to the observer. The subscript $LOS$ on the integral indicates an integral along the line of sight.  Finally, $\lambda$ indicates the wavelength of observation.  The term in square brackets is called the rotation measure (hereafter denoted by the variable RM and reported in the SI units of rad/m$^2$), and is the physical quantity retrieved in Faraday rotation measurements.

\begin{figure}
	\begin{center}
	\includegraphics[height=2.5in, trim = {25mm 48mm 25mm 40mm}, clip]{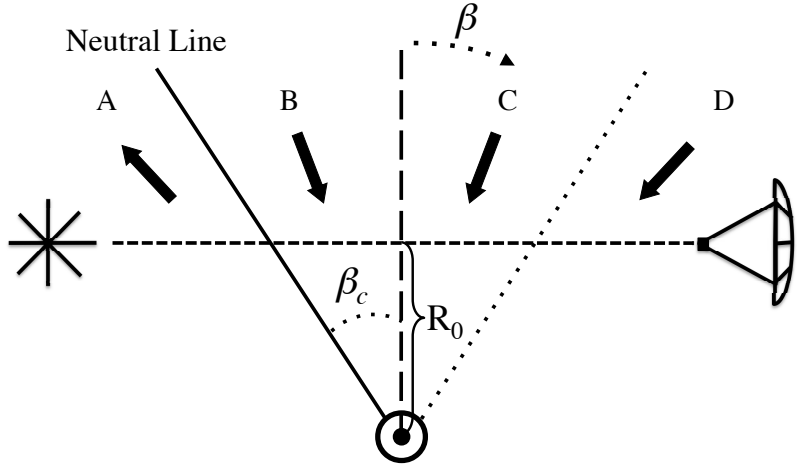}
	\caption{Illustration of the line of sight from a radio source, through the corona, to a radio telescope on Earth.  The line of sight passes at a closest distance $R_0$ which is referred to as the impact parameter.  The figure illustrates an idealization that will be employed in this paper, which is that the coronal magnetic field is radial (solid arrows).  The value of the rotation measure depends sensitively on the location of the magnetic neutral line along the line of sight (given by the angle $\beta_c$).  The solid line indicates the coronal magnetic neutral line, the dashed line divides the line of sight in two halves of equal length, and the dotted line indicates a symmetry line.  This symmetry is such that, for the case of radial magnetic field strength dependent only on heliocentric distance, the RM contributions from sectors B and C ($-\beta_c<\beta<\beta_c$) cancel.}
	\label{fig:cartoon1}
	\end{center}
\end{figure}

The geometry involved in a Faraday rotation measurement is illustrated in Figure~\ref{fig:cartoon1}. 
Of particular note for further discussion are two parameters defined in this figure. The first is the smallest heliocentric distance of any point along the line of sight, referred to as the ``impact parameter,'' and noted by $R_0$ (in units of $R_\odot$).  The second is the angle $\beta$, which is an equivalent variable to $s$ for specifying the position along the line of sight; it is defined as positive toward the observer.  As is true of many astronomical measurements, Faraday rotation yields a path-integrated measurement of the magnetic field and the plasma density.  Thus Faraday rotation is sensitive not only to the magnitude of $n_e$ and $\mathbf{B}$, but also to the orientation of $\mathbf{B}$ with respect to the line of sight.  Because this measurement depends on the portion of the magnetic field parallel (or antiparallel) to the line of sight, it is possible to measure zero Faraday rotation through the corona even when strong magnetic fields or dense plasmas are present.  A simple example would be a purely unipolar, radial magnetic field and $n_e = n_e(r)$.

Coronal Faraday rotation measurements can be made with either spacecraft transmitters or natural radio sources as the source of radio waves.  Examples of results which have been obtained with spacecraft transmitters are \cite{Levy:1969}, \cite{Stelzried:1970}, \cite{Cannon:1973}, \cite{Hollweg:1982}, \cite{Patzold:1987}, \cite{Bird&Edenhofer:1990}, \cite{Andreev:1997a}, and \cite{Jensen:2013b,Jensen:2013a}.  Observations of natural radio sources typically use either pulsars or extragalactic sources.  Observations utilizing extragalactic radio sources have been \cite{Sofue:1972}, \cite{Soboleva&Timofeeva:1983}, \cite{Sakurai&Spangler:1994a,Sakurai&Spangler:1994b}, \cite{Mancuso&Spangler:1999,Mancuso&Spangler:2000}, \cite{Spangler:2005}, \cite{Ingleby:2007}, and \cite{Mancuso&Garzelli:2013}.  Coronal observations with pulsars include \cite{Bird:1980}, \cite{Ord:2007}, and \cite{You:2012}.  Pulsars can also be used to determine the dispersion delay through the corona simultaneously with Faraday rotation, thereby providing a means of independently estimating the plasma density contribution to the rotation measure. The primary drawback is that the dispersion delay due to the corona and solar wind is small and can not be measured with sufficient accuracy for most pulsars, with the exception of some millisecond pulsars \citep{You:2012}.

One of the advantages of extragalactic radio sources relative to spacecraft transmitters and pulsars, which will be of importance in the present investigation, is that they are extended on the sky.  This permits simultaneous measurement of Faraday rotation along as many lines of sight as there are source components with sufficiently large polarized intensities.  These lines of sight will pass through different parts of the corona, and provide information on the spatial inhomogeneity of plasma density and magnetic field.  We will use the term ``differential Faraday rotation'' to refer to the difference between spatially-separated rotation measures \citep{Spangler:2005}.  Observations of extended extragalactic radio sources with imaging interferometers like the VLA are particularly well suited for measuring differential Faraday rotation.  Such observations automatically provide the RM along two or more closely-spaced lines of sight.  Obtaining this kind of information from spacecraft transmitters requires simultaneous tracking periods with two separated antennas.  Such observations have, in fact, been carried out and analyzed \citep[see, e.g.,][]{Bird:2007}.  Extragalactic radio sources, being extended, also provide a ``screen'' of polarized emission which allows examination of small scale fluctuations in the rotation measure. Finally, because extragalactic radio sources emit, and are polarized, over a wide range in radio frequency, one can test for the $\lambda^2$ dependence of polarization position angle and resolve $n \pi$ ambiguities in the position angle ($n\in\mathbb{Z}$) and insure that position angle changes measured are indeed due to Faraday rotation.  

This paper presents the results of observations of the radio galaxy 3C228 \citep{Johnson:1995} on August 17, 2011.  It was observed through the corona at impact parameters which ranged from $4.6 R_{\odot}$ to $5.0 R_{\odot}$.  Over the course of this observation, the line of sight penetrated deeper into the corona, approaching the solar north pole (see Section~\ref{sec:Geometry}).  

There are several reasons why these observations represent a significant improvement and extension of  our previous coronal Faraday rotation measurements \citep{Sakurai&Spangler:1994a,Sakurai&Spangler:1994b,Mancuso&Spangler:1999,Mancuso&Spangler:2000,Spangler:2005,Ingleby:2007}: 
\begin{enumerate}
\item Previous measurements from $1-2$ GHz of extragalactic radio sources have generally been limited to heliocentric distances $>5R_\odot$ because of a reduction in sensitivity due to solar interference in the beam side lobes \citep[for further discussion, see][]{Whiting&Spangler:2009}.  At 5.0 and 6.1 GHz, the primary beam size is small enough to allow us to make these measurements without loss of sensitivity.
\item While several observations of satellite downlink signals have been made with impact parameters of $2-5R_\odot$, these were performed at one frequency and, generally, with one line of sight.  Our observations provide two lines of sight over multiple frequencies.
\item The target source 3C228 has components which are strongly linearly polarized.  The two brightest components of the source, the northern and southern hotspots (see Section~\ref{sec:3C228}) have polarized intensities of 13 and 14 mJy/beam at 5.0 GHz in the VLA A array, relative to a typical polarized intensity map noise level of 21 $\mu$Jy/beam. 
\item The angular size of 3C228 is reasonably well suited for observations in the A array configuration, in which the VLA antennas are maximally extended.  Our experience has shown that A array observations are less susceptible to solar interference than observations with the B, C, or D arrays because there are fewer short baselines. 
\item As a consequence of the previous points, the present observations are the best to date in our research program for the goal of measuring differential coronal Faraday rotation and rotation measure fluctuations, as well as providing information for the coronal magnetic field at heliocentric distances $<5R_\odot$.
\end{enumerate}

The organization of this paper is as follows.  In Section~\ref{sec:obs}, we discuss the source characteristics of radio galaxy 3C228, the geometry of the observations, the method for data reduction, and the imaging and analysis.  In Section~\ref{sec:results}, we give our results for the mean Faraday rotation measurements over the eight hour observation period as well as the slow variations in rotation measure.  In Section~\ref{Sec:Model}, we discuss three simple, analytic models for the plasma structure and magnetic field of the corona and their associated estimates of the rotation measure and demonstrate the importance of accounting for the streamer belt.  In Section~\ref{sec:discussion}, we discuss the implications of our measurements for coronal heating by Joule and wave heating models and summarize our results and conclusions in Section~\ref{sec:Summary}.


\section{Observations and Data Analysis}\label{sec:obs}

\subsection{Properties of 3C228}\label{sec:3C228}
\begin{figure}[htb!]
	\centering
	\subfigure{
		\includegraphics[height=3.8in,trim = {50mm 0mm 80mm 0mm}, clip]{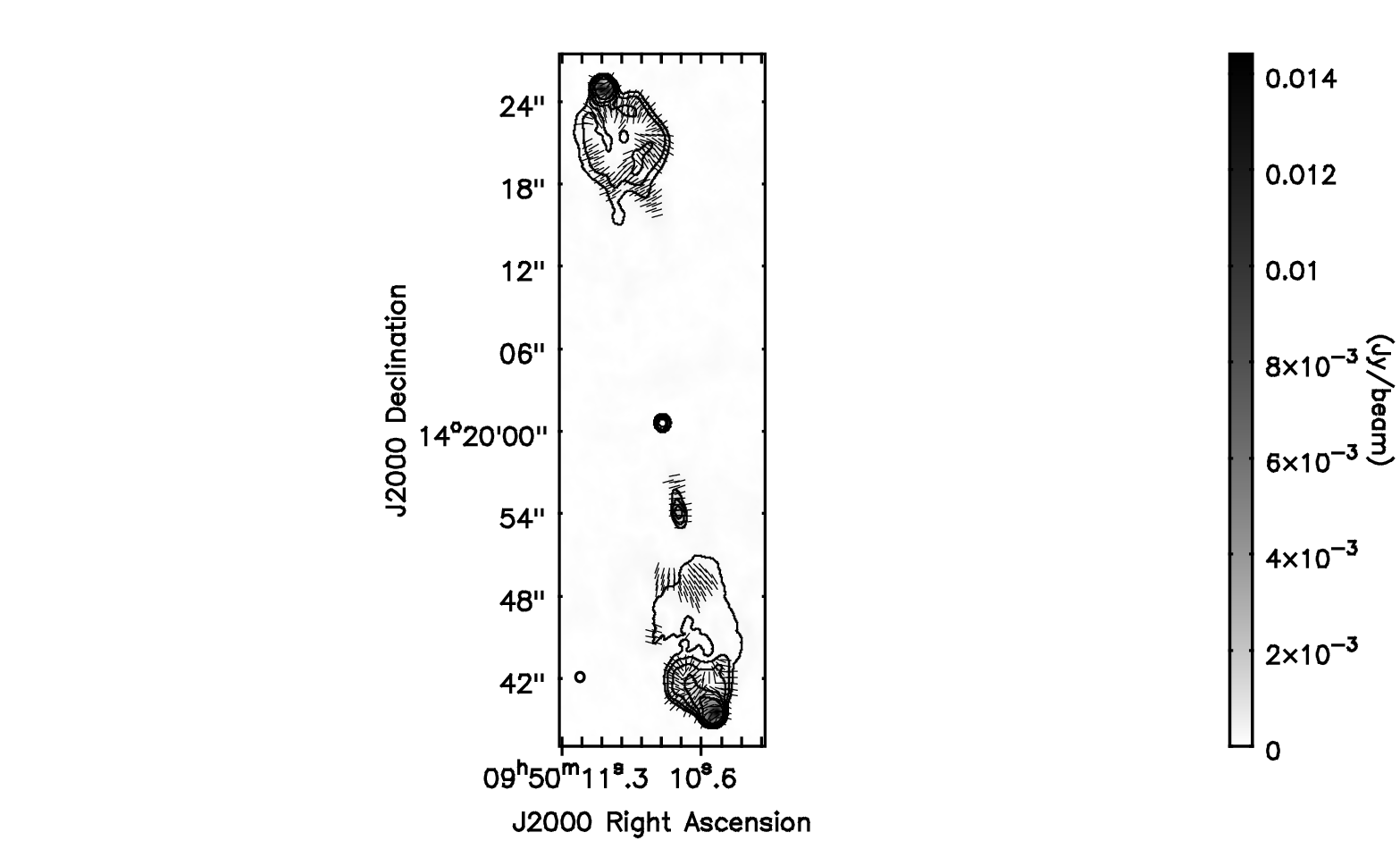}
          }
	\hspace{-.2in}
	\subfigure{
		\includegraphics[height=3.8in,trim = {50mm 0mm 80mm 0mm},clip]{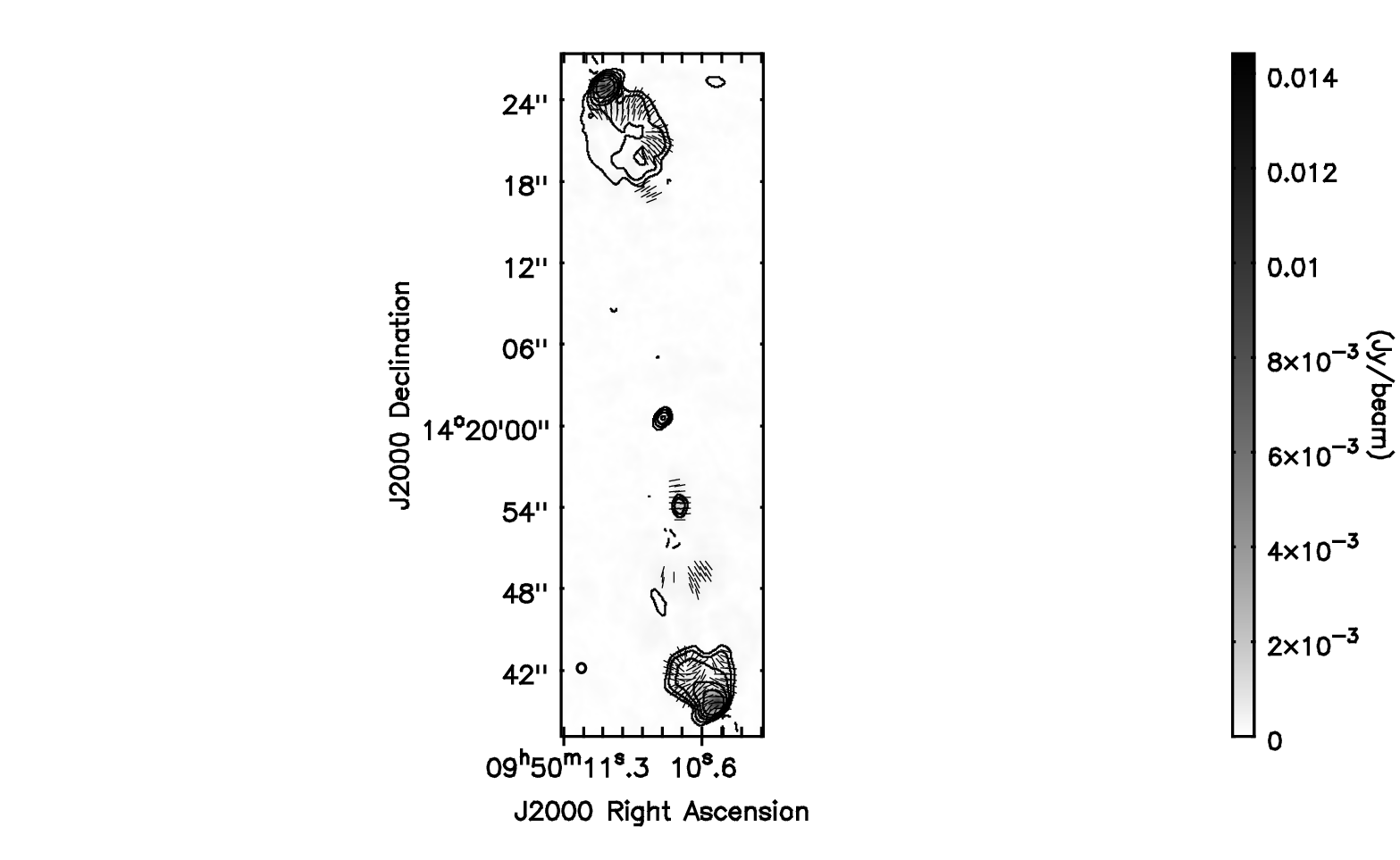}
	}
	\hspace{-.2in}
	\subfigure{
		\includegraphics[height=3.8in,trim = {170mm 0mm 0mm 0mm},clip]{f02b.eps}
	}
	\caption[]{Clean maps of the total intensity and polarization structure of the radio source 3C228 on the reference day ({\it left}) and on the day of occultation ({\it right}). These images are a synthesis of the four bandpasses centered at a frequency of 4.999 GHz using the data for the entire observation session. The contours show the distribution of total intensity (Stokes parameter I), and are plotted at -5, 5, 10, 20, 40, and 100 times the RMS noise of the total intensity ($\sigma_I = 0.207$ mJy/beam on the reference day). The peak intensity of the northern and southern hotspots was 131 and 184 mJy/beam on the reference day, respectively.  The grayscale indicates the magnitude of the polarized intensity.  The orientation of the line segments gives the polarization position angle $\chi$.  The resolution of the image (FWHM diameter of the synthesized beam) is 0.68 arcseconds.  The synthesized beam is plotted in the lower left corner of the figure.}
	 \label{fig:3C228_Maps}
\end{figure}

The basis of this paper is observations made in August, 2011, during the annual solar occultation of the radio galaxy 3C228.  The source 3C228 is a radio galaxy with a redshift of $z=0.5524$ \citep{Spinrad:1985}.  An image of this source is shown in Figure~\ref{fig:3C228_Maps}, the left image made from our VLA observations when the Sun was far from the source (see Section~\ref{sec:Observation}).

Figure \ref{fig:3C228_Maps} shows that 3C228 represents a distributed, polarized source of radio waves, ideal for probing of the corona with Faraday rotation.  The polarized emission is particularly strong in the northern hotspot at (J2000) RA = $09\tablenotemark{h}50\tablenotemark{m}11\fs1$ and DEC = $14\degr20\arcmin25\arcsec$, a jet in the southern lobe at (J2000) RA = $09\tablenotemark{h}50\tablenotemark{m}10\fs7$ and DEC = $14\degr19\arcmin54\arcsec$, and the southern hotspot at (J2000) RA = $09\tablenotemark{h}50\tablenotemark{m}10\fs5$ and DEC = $14\degr19\arcmin39\arcsec$.  The stronger components, i.e. those with the largest value of the polarized intensity, were the northern and southern hotspots, with polarized intensities $P=13 \mbox{ and } 14$ mJy/beam on the reference day, respectively.  The fractional polarization, $m=P/I$, of the northern and southern hotspots was 14\% and 8\%, respectively, at this frequency and resolution.

The peak intensity of the northern and southern hotspots was $I=131 \mbox{ and } 184$ mJy/beam on the reference day, respectively.  On the day of occultation, the peak intensity of the northern and southern hotspots decreased to $I= 93 \mbox{ and } 135$ mJy/beam, respectively; similarly, the polarized intensities decreased to $P=8 \mbox{ and } 9$ mJy/beam.  This decrease in intensity is partially due to minor angular broadening effects typically attributed to small scale coronal turbulence, which is further discussed in Section~\ref{Sec:Imaging}.

The separation between the components is important, because it determines the separation in the corona between lines of sight on which the rotation measure is determined.  The separation between the northern and southern hotspots is $46\arcsec$, corresponding to 33,000 km separation between the lines of sight in the corona.


\subsection{Geometry of the Occultation}\label{sec:Geometry}
\begin{figure}[htb!]
	\begin{center}
	\includegraphics[width=4.25in, trim = {100mm 150mm 100mm 145mm}, clip]{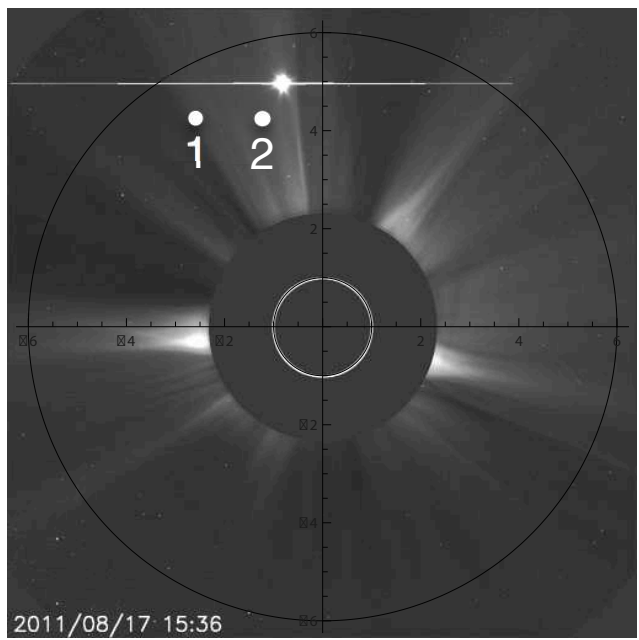}
	\caption{Solar corona on August 17, 2011, as observed with the Large Angle and Spectrometric Coronagraph (LASCO)/C2 coronagraph.  The white plotted points labeled ``1'' and ``2'' are the locations of 3C228 at the beginning (15:33 UTC) and end (23:27 UTC) of observations respectively.  The bright source north of 3C228 is Venus.  The white circle in the image center is the photosphere and the horizontal axis is the heliographic equator.  This image is from the LASCO public archive: \url{http://sohowww.nascom.nasa.gov}.}
	 \label{fig:3C228_Position}
	\end{center}
\end{figure}

The orientation of the Sun and 3C228 on August 17, 2011, is shown in Figure~\ref{fig:3C228_Position}.
During the observing session, the orientation of the line of sight changed relative to the corona.  The most important parameter describing the line of sight is the distance from the center of the Sun to the proximate point along the line of sight, termed the impact parameter $R_0$. The heliographic latitude and longitude of the proximate point are also important.  During the August 17 session (details presented in Section~\ref{sec:Observation} below), the impact parameter ranged from $5.0 - 4.6 R_\odot$ and there was a corresponding increase in the heliographic latitude of the proximate point from $59\fdg4$ to $71\fdg0$ and decrease in the Carrington longitude from $17\fdg3$ to $5\fdg8$.  As may be seen in Figure~\ref{fig:3C228_Position}, the line of sight to 3C228 primarily sampled the region over the solar north pole, where magnetic field lines are thought to be nearly unipolar and radial.


\subsection{Observations and Data Reduction}\label{sec:Observation}

All observations were performed using the Karl G. Jansky Very Large Array (VLA) of the National Radio Astronomy Observatory (NRAO)\footnote{The Karl G. Jansky Very Large Array is an instrument of the National Radio Astronomy Observatory. The National Radio Astronomy Observatory is a facility of the National Science Foundation operated under cooperative agreement by Associated Universities, Inc.} and all data reduction was performed with the Common Astronomy Software Applications (CASA) data reduction package.  Two identical sets of observations were performed: one set when the program source was occulted by the corona, and one set when the program source was distant from the Sun, allowing for measurement of the source's intrinsic polarization properties, unmodified by the corona.  The observations of the occulted program source were performed on August 17, 2011, from 15:33 to 23:27 UTC.  The reference observations were performed on August 6, 2011, from 16:16 to 00:10 UTC on August 7.  Details of the observations and resultant data are given in Table~\ref{T:Log_obs}.

The observations were similar to those previously reported in \cite{Sakurai&Spangler:1994a} and \cite{Mancuso&Spangler:1999,Mancuso&Spangler:2000}, and described in those papers.  The main features of the observations are briefly summarized below.  We also indicate features of the new observations which differ from those of our previous investigations. 
\begin{enumerate}
\item Observations were made in the A array, which provides the highest angular resolution, but more importantly, the data are less affected by solar interference.  We used an integration time of 5 seconds which, in A configuration, corresponds to an acceptable $\sim5\%$ time averaging loss in signal amplitude\footnote{See the Observational Status Summary documentation for the VLA at \url{https://science.nrao.edu/facilities/vla/docs}}.
\item Simultaneous observations were made at 4.999 GHz (6 cm) and at 6.081 GHz (5 cm), both with a 128 MHz bandwidth.  These frequencies are sufficiently separated as to allow detection of the $\lambda^2$ scaling of the Faraday rotation, an important check on the validity of the measurement.
\item Due to radio frequency interference (RFI), the upper half of the bandwidth around 6.081 GHz had to be excised as well as one antenna at 4.999 GHz and eight antennas at 6.081 GHz for both sessions.

\begin{table}[htdp]
	\centering
	\caption{Log of Observations}\label{T:Log_obs}
	\smallskip
	\begin{threeparttable}
		\begin{tabular}{lll}
		\hline\hline
		Dates of Observations												&	August 6, 2011		&  August 17, 2011\\
		Duration of Observing Sessions [h]									&	\multicolumn{2}{c}{7.89}\\
		Frequencies of Observations\tablenotemark{a} [GHz]			&	\multicolumn{2}{c}{4.999, 6.081}\\
		VLA array configuration												&	\multicolumn{2}{c}{A}\\
		Restoring Beam (FWHM)\tablenotemark{b}						&	\multicolumn{2}{c}{0\farcs68\tablenotemark{c}}\\
		Range in Solar Impact Parameter, $R_0$ [$R_\odot$]			&	$41.4-42.6$		&	$4.6-5.0$\\
		RMS noise level in I Maps\tablenotemark{b} [mJy/beam]		&	0.207, 0.333\tablenotemark{d}		&	 0.176, 0.225\tablenotemark{d}\\
		RMS noise level in P Maps\tablenotemark{b} [$\mu$Jy/beam]	&	21, 45\tablenotemark{d}				&	 23, 42\tablenotemark{d,e}\\
		
		\hline
		\end{tabular}
	\medskip
		\begin{tablenotes}
		\small
		\item[a] The spectral windows centered at each frequency had a 128 MHz bandwidth.
		\item[b] This is for the maps using the data from the entire observation session.
		\item[c] The restoring beam on the day of occultation was fixed to be the same as on the reference day.
		\item[d] Average RMS levels for 4.999 and 6.081 GHz maps respectively.  
		\item[e] After two iterations of phase-only self-calibration, the ratio of peak intensity to the RMS noise (termed the dynamic range) improved by a factor of 8.5 and 3.5 for 4.999 and 6.081 GHz maps respectively.
		\end{tablenotes}	
	\end{threeparttable}
\end{table}

\item Observations were made using frequencies near 5 GHz because we were interested in performing sensitive observations close to the Sun.  Previous results reported by \citet{Sakurai&Spangler:1994a}, \citet{Mancuso&Spangler:1999}, \citet{Spangler:2005}, and \citet{Ingleby:2007} used frequencies near 1.465 and 1.635 GHz and, consequently, were limited in how close to the Sun they could make reliable measurements.  As the target sources move closer to the solar limb, solar interference enters the side lobes of the antenna beam.  This causes significant increases in the system temperature and thereby lowers the sensitivity of the measurements.  The beam size is inversely proportional to the observational frequency.  Because the beam size is smaller near 5 GHz, we can observe a natural radio source much closer to the solar limb than any previous work.  \cite{Whiting&Spangler:2009} demonstrated that the system temperatures in the upgraded VLA are manageable for 5 GHz, only increasing from a nominal value of $\sim40$ K to $50-60$ K for observations with a proximate point of $\sim4R_\odot$ during solar minimum. 
\item Observations of 3C228 were made in scans of 9 minutes in duration, bracketed by 2 minute observations of a phase calibrator, for a total of 33 scans.  The average interval between each scan was $\sim14$ minutes.
\item The main calibrator for both sessions was J0956+2515.  This source was used for phase and amplitude calibration, as well as measurement of instrumental polarization.  Observations were also made of another calibrator source, J0854+2006, as an independent check of the polarimeter calibration.  We found that the polarization calibration values for both sources were in excellent agreement.  The angular separation from the target source is $11\fdg0$ and $14\fdg4$ for J0956+2515 and J0854+2006, respectively.  On the day of occultation, the impact parameters for J0956+2515 and J0854+2006 were $R_0\sim46$ and $R_0\sim54$; consequently, coronal influence on the calibration scans is negligible.
\item Polarization data were not corrected for estimated ionospheric Faraday rotation.  Currently, the Common Astronomy Software Applications (CASA) data reduction package does not support ionospheric Faraday rotation corrections.  To estimate the ionospheric Faraday rotation during both sessions, we loaded the data into the AIPS (Astronomical Image Processing System) program and used the AIPS procedure VLBATECR.  The estimates of the ionospheric Faraday rotation measure ranged from $1.5-1.9$ rad/m$^2$ on August 6 and about $1.5-2.3$ rad/m$^2$ on August 17.  These are similar in part because we observed at the same local sidereal time (LST) on both days.  Because of the method used to determine the coronal Faraday rotation (see Section~\ref{Sec:Imaging}), the total contribution from ionospheric Faraday rotation is expected to be $\lesssim 1$ rad/m$^2$.
\item The instrumental polarization, described by the antenna-specific $D$ factors \citep{Bignell:1982,Sakurai&Spangler:1994b} was determined from the observations of J0956+2515 in both sessions.  The same reference antenna was used in both sessions, allowing for comparison of the $D$ factors; the amplitudes and phases of the $D$ factors were nearly identical for all antennas.  Similarly, a completely independent measurement of the $D$ factors was made with the data of J0854+2006 for both sessions.  The independent measurements of the $D$ factors showed agreement in both amplitude and phase which was similar to that reported by \cite{Sakurai&Spangler:1994b} and illustrated in Figure 2 of that paper.  One feature of this analysis to note is that the amplitudes of the $D$ factors are higher for the upgraded VLA antennas, $D\approx 5-10\%$, than for the pre-upgrade antennas, $D\approx 1-4\%$, studied by \cite{Sakurai&Spangler:1994b}.  This difference is understood and expected.
\item The offset to the origin of the polarization position angle, expressed by the net RL phase difference, was measured by observations of 3C286.  A second calibrator with known polarization, 3C138, was observed to provide a check on the integrity of the calibration solutions determined from 3C286.  3C138 was calibrated using the RL phase difference solutions from 3C286; the measured position angle was within $0\fdg5$ of the value listed in VLA calibrator catalogs for both sessions and all observing frequencies.
\item The mean position angles for J0956+2515 agree to within $2\fdg0$ and $4\fdg6$ at 4.999 GHz and 6.081 GHz, respectively, for both sessions. This is important for establishing that there is no significant offset in the origin of the position angle between the reference session and the day of occultation.  This is further discussed in Section~\ref{sec:results_mean_rot}.
\item During the observations on the day of occultation, we monitored the system temperature as the antennas pointed closer to the Sun.  Data were taken from two antennas and gave essentially the same results: the system temperature was enhanced, with an approximate range of 1.3 to 2.3.  This enhancement is roughly consistent with, but somewhat higher than that indicated by \cite{Whiting&Spangler:2009}.  Part of this difference can be attributed to the increase in solar flux density; the observations of \cite{Whiting&Spangler:2009} were made during solar minimum while the observations reported here are during the approach to solar maximum.  For more details, see Appendix~\ref{Sec:Tsys_discussion}.
\end{enumerate}


\subsection{Imaging}\label{Sec:Imaging}
The calibrated VLA visibility data were split into 6 bandpasses with corresponding sky frequencies of 4.957 GHz (BW 28 MHz), 4.985 GHz (BW 28 MHz), 5.013 GHz (BW 28 MHz), 5.041 GHz (BW 28 MHz), 6.067 GHz (BW 28 MHz), 6.094 GHz (BW 26 MHz).  There were two reasons for this.  The first was to reduce the effects of bandwidth smearing.  The second was to allow for more rigorous validation that changes in position angle were a result of Faraday rotation.  These data were then used to generate maps using the CASA task CLEAN using the multi-frequency synthesis mode with a cell size of $0.07\arcsec$.  Because we are primarily concerned with sensitivity and not resolution, the maps were generated using a Natural Weighting scheme with a Gaussian taper of 250 k$\lambda$.  To enable accurate comparison at the 6 intermediate frequencies for each source, the maps produced from observations on August 17 were restored using the same beam size as maps from the reference observations on August 6.  Further, the 4.957 GHz beam was used to restore the 6.067 GHz and 6.094 GHz data on both days to have the same resolutions.  To produce the final set of maps, two iterations of phase-only self-calibration were performed.  This improved the ratio of peak intensity to the RMS noise (termed the dynamic range) by a factor of 8.5 and 3.5 for the four bandpasses centered at 4.999 and the two bandpasses centered at 6.081 GHz, respectively.

For each source, we generated maps in the Stokes parameters $I$, $Q$, $U$, and $V$ for each scan as well as a ``session map'' made from all the data at a given frequency.  The session maps provide a measure of the Faraday rotation averaged over the entire eight hour observing session; the individual scan maps, however, allow for examination of the time variations over the observing session, with a resolution on the order of the interval between scans: $\sim14$ minutes.  We then generated maps of the (linear) polarized intensity, $P$, and the polarization position angle, $\chi$, from these Stokes parameters:
\begin{eqnarray}
P &=& \sqrt{Q^2+U^2}\\
\chi &=& \frac{1}{2}\arctan\left(\frac{U}{Q}\right)
\end{eqnarray}
Figure~\ref{fig:3C228_Maps} shows the $I$, $P$, and $\chi$ session maps for the program source for the full eight hours on the reference day (left image) and the day of coronal occultation (right image).  $P$ is shown as a grayscale map, the contours show the distribution of $I$, and the orientation of the lines gives $\chi$.  All of these parameters vary with position in the image.  

A striking feature of these maps is the lack of visible angular broadening of the source.  This phenomenon results from density irregularities elongated along the coronal magnetic field that cause radio wave scattering \citep[see][]{Armstrong:1990,Coles:1995,Grall:1997}.  Pronounced angular broadening of 3C228 was observed during the 1.465 GHz observations in August, 2003, reported in \cite{Spangler:2005}.  While the angular broadening in the maps is not readily apparent, there is a measurable decrease in $I$ and $P$ on the day of occultation.  

In all radioastronomical observations, the measured intensity is the convolution of the true intensity with a point spread function.  In most interferometer applications, the point spread function is simply the synthesized beam, and comparison of the reference day maps with the maps on the day of occultation only requires observations with the same array.  In the present application, however, the point spread function for the program observations is the convolution of the synthesized beam with the power pattern of the angular broadening.

To determine the extent of the angular broadening, the following procedure was employed for the session maps:  
\begin{enumerate}
\item Using the reference $I$ map of August 6, two dimensional Gaussian fits were made to the northern and southern hotspots, as well as the central component at (J2000) RA = $09\tablenotemark{h}50\tablenotemark{m}10\fs8$ and DEC = $14\degr20\arcmin01\arcsec$, using the CASA task IMFIT.  This provides the Gaussian equivalent diameters of the components.
\item The known synthesized beam sizes were quadratically subtracted to obtain Gaussian equivalent, estimated angular sizes of the components.  These sizes were approximately $0.5\arcsec$ for the northern and southern hotspots and the central component was consistent with a point source (i.e., its Gaussian equivalent diameters were consistent with the beam size). 
\item Using the occultation $I$ map of August 17, two dimensional Gaussian fits were made to the northern and southern hotspots, as well as the central component.
\item The beam sizes and the estimated angular sizes of the components were quadratically subtracted to obtain the Gaussian equivalent angular broadening due to the corona.  The three components were consistent with each other, yielding a small elliptical broadening disk of $0.4\arcsec\times0.2\arcsec$ with position angle $\sim45\degr$.
\end{enumerate}

This broadening disk is consistent with a $12-18\%$ drop in $I$.  Consequently, we can account for half of the drop in intensity on the day of occultation and, similarly, half of the drop in the RMS noise level of $I$ (Table~\ref{T:Log_obs}).  While the RMS noise level of $P$ in Table~\ref{T:Log_obs} is similar on both days, the polarized intensity $P$ does decrease by $\sim40\%$ and, consequently, the dynamic range on the day of occultation decreases.

The small size of the Gaussian disk is also consistent with the lack of visible broadening in the maps (e.g., Figure~\ref{fig:3C228_Maps}); a point-like source, such as the central component, which has the same structure as the synthesized beam on the reference day, will only broaden by a factor of $5-16\%$.  Extended structures are even less affected; the northern and southern hotspots only broaden by a factor of $2-9\%$.  Because the broadening is not significant, we did not correct for this phenomenon (e.g., by convolving the session and scan maps on the reference day with an equivalent $0.4\arcsec\times0.2\arcsec$ Gaussian disk).  

We examined the session map for the reference day (left image, Figure~\ref{fig:3C228_Maps}) for local maxima in polarization intensity, typically $P>5$ mJy/beam.  We chose these locations in order to maximize the sensitivity of our measurement.  Propagation of errors yields the following expressions for the error in polarized intensity, $\sigma_P$ and, more importantly, polarization position angle, $\sigma_\chi$:
\begin{eqnarray}
\sigma^2_P &=& \left(Q^2\sigma^2_Q+U^2\sigma^2_U\right)/P^2\label{eq:sigP}\\
\sigma^2_\chi &=& \left(Q^2\sigma^2_U+U^2\sigma^2_Q\right)/4P^4\label{eq:sigX}
\end{eqnarray}
assuming the errors in $Q$ and $U$ are uncorrelated.  Typically, $\sigma_Q\approx\sigma_U$ (although this was not strictly an equality for most of our $Q$ and $U$ maps), so $\sigma_P\approx\sigma_Q\approx\sigma_U$ and $\sigma_\chi\approx\sigma_P/2P$.  Consequently, stronger $P$ provides a more robust measurement for $\chi$.  The value of $P$ was too low to allow an accurate polarization measurement over most of the source.  Even in places where polarized emission was detected over an extended area (e.g., the southern jet), the errors were too large to allow useful results on Faraday rotation.  In the rest of the paper, the analysis is based on measurements at the points in the northern and southern hotspots where the polarized intensity was at a local maximum in the session map for the reference day.  The actual location of the maximum $P$ for both hotspots did not vary significantly from this value from scan to scan.  The largest deviations were $<0.35\arcsec$ or  $\sim$ HWHM of the beam, but most deviations were $\sim0.14\arcsec$.  

We measured the values of the polarization quantities $I$, $Q$, and $U$ over a square patch consisting of 25 pixels and centered on the pixel with peak $P$.  The side of this square patch was equal to the HWHM of the beam.  We calculated the unweighted average of these measurements and then used the average to derive the polarization quantities $P$ and $\chi$ for the individual scan and session maps on both observation days.  We calculated the coronal Faraday rotation, $\Delta\chi^i(x,y)$, for the $i$th scan map by straight subtraction:
\begin{eqnarray}\label{eq:RM_calc_diff}
\Delta\chi^i(x,y) &=& \chi^i_{occ}(x,y) - \chi^i_{ref}(x,y)
\end{eqnarray}
for $i\in[1,33]$, where $\chi^i_{occ}(x,y)$ and $\chi^i_{ref}(x,y)$ are the polarization position angles at a location $(x,y)$ on the day of occultation and reference observations, respectively.

We employed this method of subtracting the individual scan maps in order to eliminate Faraday rotation caused by the background interstellar medium and reduce the effects of polarimeter calibration error, which would typically require second order instrumental polarization calibration \cite[e.g., see][]{Sakurai&Spangler:1994b}.  Further, because we performed the observations at the same LST for both days, this subtraction method reduces the contribution from ionospheric Faraday rotation (see Section~\ref{sec:results_slow_rot}).  The ``mean'' rotation, $\overline{\Delta\chi}$, for the full eight hour session was calculated in the same fashion.  Having calculated $\Delta\chi$ for each of the 6 bandpasses, we then used a least squares algorithm to determine the rotation measure, RM, for each individual scan as well as a mean $\overline{\rm{RM}}$ for the eight hour session using Equation~\eqref{eq:FR}.  We used a fit weighted by the radiometer noise because the fidelity of the data in the 4 bandpasses centered at 4.999 GHz is much better than the data in the 2 bandpasses centered at 6.081 GHz due to RFI (see Section~\ref{sec:Observation}).


\section{Observational Results}\label{sec:results}

\subsection{Mean Rotation Measure}\label{sec:results_mean_rot}

\begin{figure}[htb!]
	\begin{center}
	\includegraphics[height=2.5in, trim = {15mm 110mm 10mm 0mm}, clip]{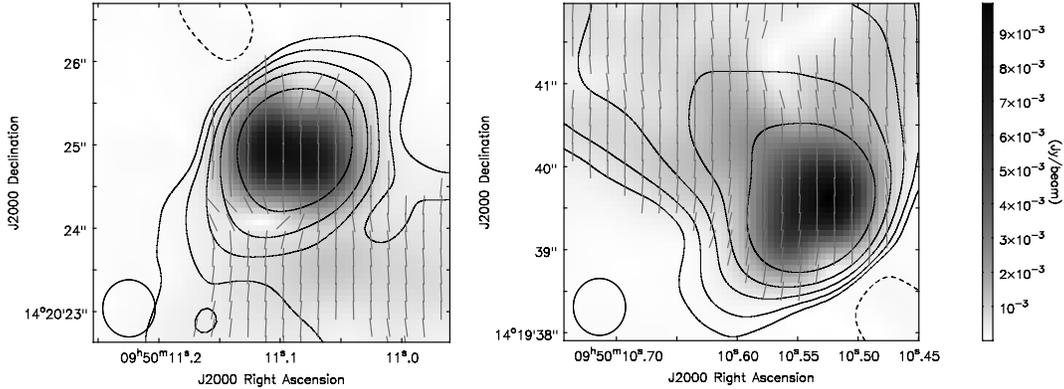}
	\caption{Clean maps of the total intensity and polarization structure of the radio source 3C228 for the northern hotspot ({\it left}) and southern hotspot ({\it right}). These images are a synthesis of the four bandpasses centered at a frequency of 4.999 GHz using the data for the entire observation session. The contours show the distribution of total intensity (Stokes parameter I), and are plotted at -5, 5, 10, 20, 40, and 100 times the RMS noise of the total intensity ($\sigma_I = 0.207$ mJy/beam on the reference day).  The grayscale indicates the magnitude of the polarized intensity.  The orientation of the lines gives the difference, or rotation, in the polarization position angle $\overline{\Delta\chi}$.  The resolution of the image (FWHM diameter of the synthesized beam) is 0.68 arcseconds. The synthesized beam is plotted in the lower left corner of the figure.
	}
	 \label{fig:3C228_Component_Maps}
	\end{center}
\end{figure}

The polarization structure for the northern and southern hotspots on the day of coronal occultation is shown in Figure~\ref{fig:3C228_Component_Maps}.  These maps are the same as the right image in Figure~\ref{fig:3C228_Maps} except that the orientation of the lines corresponds to $\overline{\Delta\chi}$, the mean rotation over the entire session of observation.  
The most striking feature is that $\overline{\Delta\chi}\sim0\degr$.  From the results of previous work \cite[e.g.,][]{Mancuso&Spangler:1999,Spangler:2005,Ingleby:2007}, we expected to see a large $\overline{\Delta\chi}$, on the order of $10\degr -20\degr$ (see Section~\ref{Sec:Model_1}). Figure~\ref{fig:3C228_Component_Maps} also demonstrates that $\overline{\Delta\chi}$ is small over most of the polarized intensity map, not just where $P$ is strongest.

We report the mean Faraday rotation, $\overline{\Delta\chi}$, for 4.999 GHz and 6.081 GHz and the mean rotation measure, $\overline{\rm{RM}}$, determined from averaging over the HWHM of the beam centered at the pixel of maximum polarized intensity in Table~\ref{T:Mean_RM}.  
\begin{table}[htdp]
	\begin{center}
	\caption{Mean Rotation Measures of 3C228}\label{T:Mean_RM}
	\smallskip
	\begin{threeparttable}
		\begin{tabular}{lccc}
		\hline\hline
		Source				&	Frequency [GHz]		&	$\overline{\Delta\chi}$ [degrees]	&	$\overline{\rm{RM}}$\tablenotemark{a} [rad/m$^2$]	\\
		\hline	
		Northern Hotspot		&	4.999					&	$-0.08\pm0.07$\tablenotemark{b}			&	$-0.69\pm0.33$\\
								&	6.081					&	$-0.65\pm0.17$\tablenotemark{c}		&	\\
		Southern Hotspot		&	4.999					&	$1.06\pm0.07$\tablenotemark{b}			&	$4.83\pm0.31$\\	
								&	6.081					&	$0.15\pm0.16$\tablenotemark{c}			&	\\	
		\hline
		\end{tabular}
	\medskip
		\begin{tablenotes}
		\small
		\item[a] Weighted average of all 6 bandpasses.
		\item[b] Weighted average of the 4 bandpasses centered at 4.999 GHz.
		\item[c] Weighted average of the 2 bandpasses centered at 6.081 GHz.
		\end{tablenotes}	
	\end{threeparttable}
	\end{center}
\end{table}
Consequently, the 25 pixels used for this measurement all lie within the $100\sigma_I$ contour in Figure~\ref{fig:3C228_Component_Maps} and only include the region where polarized emission is strongest.  While the mean position angles for the phase calibrator agree to within $\sim2\fdg0$ and $\sim4\fdg6$ at 4.999 GHz and 6.081 GHz, respectively, for the two observing sessions, there is no such systematic offset in the values presented in Table~\ref{T:Mean_RM}. Consequently, even though there is an inconsistency in the mean polarization position angles for the phase calibrator J0956+2515, this does not appear to affect our target source.

In general, the errors in measuring the RM for the northern hotspot are similar to those for the southern hotspot for the total session as well as the individual scans for a given frequency.  The errors for observations at 6.081 GHz are typically $2.4-3.5$ times greater than for observations at 4.999 GHz.  As discussed in Section~\ref{sec:Observation}, half of the bandwidth centered at 6.081 GHz had to be excised as well as eight antennas due to RFI during both observing sessions.  This alone accounts for a decrease in sensitivity by 2.1, which is consistent with the decrease in sensitivity demonstrated in Table~\ref{T:Mean_RM}.

The $\overline{\rm{RM}}$ for the northern hotspot is consistent with zero, while the southern hotspot yields $4.83$ rad/m$^2$.  
Values of $|\text{RM}|\approx 5$ rad/m$^2$ are reasonable at heliocentric distances of $\sim10R_\odot$ by standards of previous observations; however, these values are lower than would be expected at shorter distances (e.g., see Section~\ref{Sec:Model}).  The fact that the values for the northern and southern hotspots are not fully consistent with each other may be indicative of the presence of differential Faraday rotation.


\subsection{Slowly Varying Rotation Measure}\label{sec:results_slow_rot}

\begin{figure}[htb!]
	\begin{center}
	\includegraphics[width=4.25in]{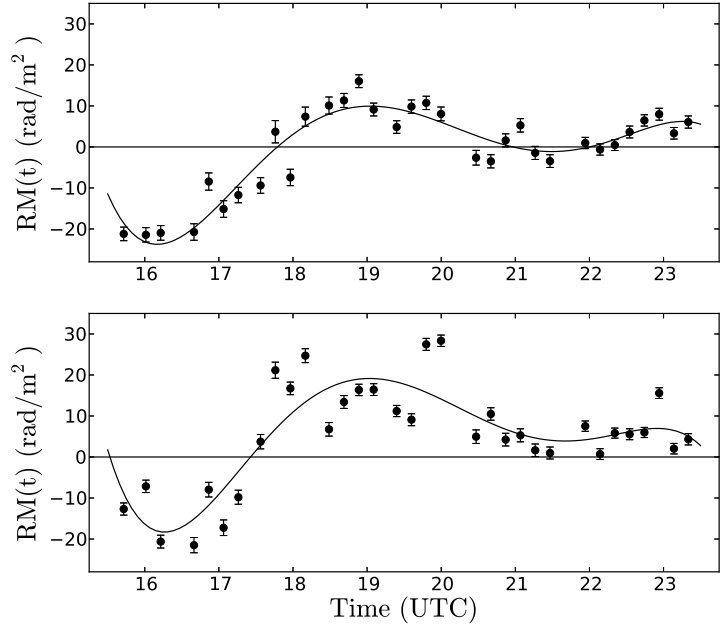}
	\caption{Coronal Faraday rotation RM(t) to 3C228 on August 17, 2011, for the northern hotspot ({\it top}) and the southern hotspot ({\it bottom}).  Each point represents the RM determined from all 6 bandpasses for a given scan ($\sim9$ minutes duration).  As time increases, the solar impact parameter $R_0$ decreases from $5.00 R_\odot$ at 15:42 UTC to $4.58 R_\odot$ at 23:19 UTC.  The superposed curve is a fifth-order polynomial fit to the data.}
	 \label{fig:RM_Time_Series}
	\end{center}
\end{figure}

The data for the rotation measure on a scan-by-scan basis show that the RM was not constant, but changed gradually over time during the observation session.  The time series of coronal Faraday rotation, RM($t$), is shown in Figure~\ref{fig:RM_Time_Series} together with a fifth-order polynomial fit that represents the slowly varying rotation measure profile.  
For both components, the rotation measure was $\approx -20$ rad/m$^2$ at the start of the observing session ($R_0 = 5.0$), approached a maximum of $\approx +20$ rad/m$^2$ ($R_0 = 4.8$), and slowly decreased to $\sim0-5$ rad/m$^2$ near the end of the observing session ($R_0 = 4.6$).  The average of the individual scan values in this figure is $-0.64\pm0.30$ rad/m$^2$ for the northern hotspot and $5.26\pm0.27$ rad/m$^2$ for the southern hotspot; these values are consistent with the mean rotation measures, $\overline{\rm{RM}}$, in Table~\ref{T:Mean_RM}.

Here, we address the issue of ionospheric Faraday rotation mentioned in Section~\ref{sec:Observation}.  Although we can not correct for ionospheric Faraday rotation in CASA, we can use the AIPS procedure VLBATECR to estimate the magnitude and sign of this effect (see Section~\ref{sec:Observation}).  We can approximate this effect by subtracting the contribution of the ionospheric Faraday rotation determined for an individual scan on the reference day from the contribution for the corresponding scan on the day of occultation (i.e., Equation~\eqref{eq:RM_calc_diff}).  Doing so, we find that the total ionospheric contribution to the rotation measures in Figure~\ref{fig:RM_Time_Series} is expected to be $\lesssim1$ rad/m$^2$, which is less than the error introduced by radiometer noise.  Consequently, we conclude these trends are largely unaffected by ionospheric Faraday rotation.

\begin{figure}[htb!]
	\begin{center}
	\includegraphics[width=4.25in]{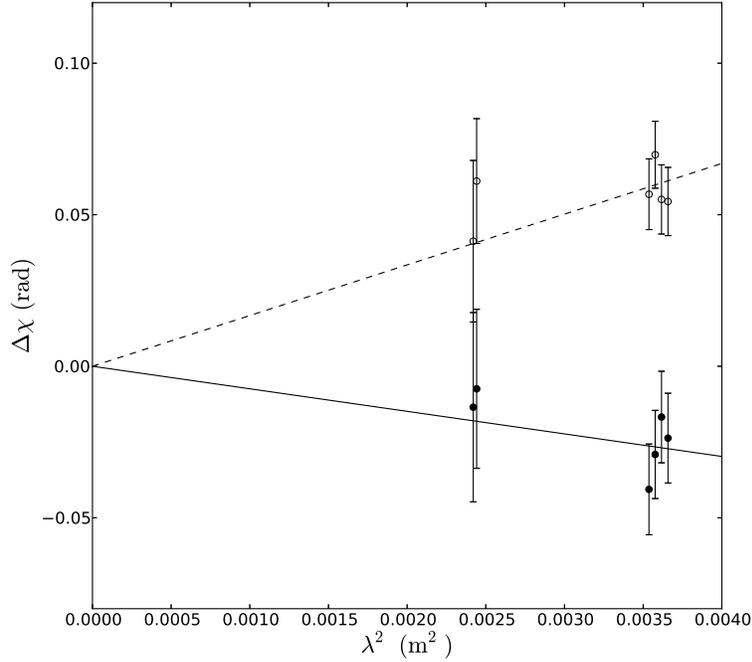}
	\caption{Polarization position angle rotation, $\Delta\chi$, as a function of the square of the wavelength, $\lambda^2$, for 3C228 at 17:57 UTC.  The solid and empty circles correspond to measurements of the northern and southern hotspots, respectively.  The solid and dashed lines are weighted least squares fits to the 6 bandpasses for the northern and southern hotspots, respectively.  The slope of the solid and dashed lines are RM $= -7.4\pm2.0$ rad/m$^2$ and RM $= 16.7\pm1.5$  rad/m$^2$, respectively.  Differential Faraday rotation is clearly present.
	}
	 \label{fig:DELTA_X_Scan_10}
	\end{center}
\end{figure}

It is constructive to examine a typical plot of the rotation in polarization position angle, $\Delta\chi$, as a function of the square of the wavelength, $\lambda^2$.  Figure~\ref{fig:DELTA_X_Scan_10} shows these data separately for the northern and southern hotspots of 3C228 at 17:57 UTC.  As discussed in Section~\ref{sec:results_mean_rot} and also demonstrated in Table~\ref{T:Mean_RM}, there is no systematic offset in the four bandpasses centered at 4.999 GHz nor is there a systematic offset in the two bandpasses centered at 6.081 GHz as may be introduced by errors during phase calibration.  Further, the 6 bandpasses are consistent with each other; the expected $\lambda^2$ dependence is clearly present.  We therefore believe the rotation measure profiles in Figure~\ref{fig:RM_Time_Series} are not a consequence of systematic effects introduced by the errors in the phase calibrator mentioned in Section~\ref{sec:Observation}.  Figure~\ref{fig:DELTA_X_Scan_10} also demonstrates the reality of the large RM values in Figure~\ref{fig:RM_Time_Series}, and is an example of the differential Faraday rotation present during our observations.

From the above arguments, we conclude that the trends in the RM$(t)$ time series must result from  the coronal plasma and not the ionospheric plasma or instrumental defects.  Consequently, we interpret the smooth changes shown in Figure~\ref{fig:RM_Time_Series} (which are on timescales of several hours) as a result of the line of sight moving past quasi-static plasma structures in the corona.  The most striking features of these observations is the s-shape of the slow time variations and the similarity in the mean profiles.  In previous work that examined slow time variations in Faraday rotation \citep{Mancuso&Spangler:1999,Spangler:2005}, the absolute magnitude of the rotation measure slowly increased in time (albeit, non-monotonically), which was believed to be a result of deeper penetration of the corona by the line of sight.  Our results are surprising in that as the line of sight increases in proximity to the Sun, the Faraday rotation approaches $\sim0-5$ rad/m$^2$.

The s-shaped trend can be understood at least qualitatively by examining LASCO/C2 coronagraph images taken during the observing session.  From the beginning of the observations, 15:33 UTC, to about 20:00 UTC, the line of sight was occulted by a bright coronal ray off the northeastern limb of the Sun.  Further, two coronal mass ejections were observed on this day, in particular one originating on the northeastern limb passed through the region sampled by the line of sight to 3C228 between 4:00 and 10:00 UTC, just hours before our observations.  There are also apparent changes in brightness to the coronal ray we attribute to asymmetric plasma flow continuing from 10:00 to 14:00 UTC, then minor changes in brightness from 14:00 to 17:30 UTC.  Consequently, we believe the magnetic field lines and plasma density profile in this region were more complex during the beginning of our observations and the change in sign for RM$(t)$ in Figure~\ref{fig:RM_Time_Series} results from sampling regions of different magnetic orientations and densities.

The decrease in magnitude of RM$(t)$ near the end of the observing session is likely a consequence of the geometry of the line of sight.  Examination of the LASCO/C2 coronagraph images reveals that the line of sight enters into a coronal hole after 20:00 UTC.  Thus we expect a reduction in RM$(t)$ due to the reduced plasma density in this region.  Further, Figure~\ref{fig:3C228_Position} shows that the line of sight to 3C228 approaches the northern solar pole near the end of the observing session.  The polar regions are well-known to contain open, unipolar field lines and, consequently, the polar regions have appreciably more symmetric magnetic field lines than the equatorial regions.  Because the rotation measure is the sum of all $n_eB_\parallel$ contributions along the line of sight (Equation~\eqref{eq:FR}), if the line of sight passes through a region with an increasingly symmetric magnetic field, the contributions will sum to zero for $n_e=n_e(r)$.  Had the line of sight to 3C228 approached the equatorial region instead, the measured Faraday rotation would almost certainly have been much greater.

In Figure~\ref{fig:RM_Time_Series}, the mean profiles appear to be quite similar.  While the scans for the northern hotspot track the mean profile well, there is more variation for the southern hotspot.  The similarity in the mean profiles raises the question of whether or not the difference between $\overline{\rm{RM}}$ for the northern and southern hotspots is (1) a consequence of the fluctuations primarily present in the southern hotspot as may be associated with periods of differential Faraday rotation (see Section~\ref{sec:discuss_Diff_Far}) or (2) a result of fundamentally different mean profiles.  Removing the five scans with the largest fluctuations about the mean profile (for both components) and recalculating the average of the remaining scans yields $-0.61\pm0.33$ rad/m$^2$ and $2.32\pm0.29$ rad/m$^2$ for the northern and southern hotspots, respectively.  $\overline{\rm{RM}}$ for the northern hotspot is largely unaffected by this procedure, but $\overline{\rm{RM}}$ for the southern hotspot decreases $>50\%$.  This indicates that the difference is partially a consequence of the fluctuations primarily present in the southern hotspot.

Because of the coincidence between the profile of the RM($t$) and the session mean $\overline{\rm{RM}}\sim0$ rad/m$^2$ for the northern hotspot and $\overline{\rm{RM}}\sim5$ rad/m$^2$ for the southern hotspot, a second set of maps of 3C228 were made in which three or four scans were averaged together to increase integration time and $(u,v)$ plane coverage.  These maps did not reveal any qualitatively or quantitatively new results with respect to the individual scan measurements described above.


\section{Comparison of Observations with Model Coronas}\label{Sec:Model}

The principal importance of Faraday rotation measurements is the information they provide on the strength and functional form of the coronal magnetic field in the heliocentric distance range $4 R_{\odot} \leq r \leq 10 R_{\odot}$.  Equation~\eqref{eq:FR} shows that, to obtain information on the magnetic field, a model for the plasma density $n_e$ needs to be supplied.  An initial model for the structure of $\mathbf{B}$ also should be supplied, to indicate regions of differing polarity along the line of sight.  Both models can then be iterated to achieve agreement with observations.  

In this section, we will employ simplified analytic expressions for the coronal plasma density and magnetic field as we have done in prior investigations.   These expressions allow closed form, analytic expressions for the rotation measure along a given line of sight which permit easy assessment of the agreement between models and observation, and identification of the sources of disagreement.  The three models discussed in Sections~\ref{Sec:Model_1},~\ref{Sec:Model_2}, and~\ref{Sec:Model_3} correspond to increasing complexity and realism in modeling the coronal plasma.


\subsection{Analytic Models of the Corona}\label{Sec:Model_sec}

\subsubsection{Model 1 - Uniform Corona with Single Power Laws in Density and Magnetic Field}\label{Sec:Model_1} 
Consider the coronal geometry shown in Figure~\ref{fig:cartoon1}.  The position along the line of sight is indicated by the angle $\beta$.  As shown in \cite{Patzold:1987}, use of this variable allows simplification of the integral in Equation~\eqref{eq:FR}.  In our model, we assume that the plasma density depends only on the heliocentric distance, and that the magnetic field is entirely radial \cite[a good approximation at the distances of interest in this project, see, e.g.,][]{Banaszkiewicz:1998}, and that its magnitude depends only on heliocentric distance $r$.  The coronal current sheet (here, an infinitely thin neutral line), where the polarity of the coronal magnetic field reverses, is located at an angle $\beta_c$.  

In this drawing, the line of sight crosses only one neutral line.  This was the case for our August, 2011 observations.  From symmetry considerations, it can be seen that the contributions to the integral in Equation~\eqref{eq:FR} from zones B \& C cancel, while those of A \& D make equal contributions of the same sign.  

As may be appreciated from this diagram, as well as the equations presented below, the angle $\beta_c$ is crucial in determining the magnitude of the rotation measure.  To determine $\beta_c$, the following procedure was followed. First, a program was used which projected the line of sight onto heliographic coordinates.  Second, maps of the potential field estimate of the coronal magnetic field at $r=3.25 R_{\odot}$ were obtained from the online archive of the Wilcox Solar Observatory (WSO).  The digital form of these maps was used to determine the heliographic coordinates of the coronal neutral line.  The value of $\beta$ at which these two curves intersected gave the parameter $\beta_c$.  For more details, see \cite{Mancuso&Spangler:2000} and \cite{Ingleby:2007}.  We determined $\beta_c$ using values for the two Carrington rotations that had data for August 17, 2011: 2113 and 2114, which gave $\beta_c \approx 33^\circ$ and $\approx42^\circ$ respectively.  The real value likely lies within this spread of $\beta_c$.

The specific forms for the single power law representations of the plasma density and magnetic field were as follows: 
\begin{eqnarray}
n_e(r) = N_0 \left(\frac{r}{R_{\odot}} \right)^{-\alpha}\label{eq:single_ne}\\
\mathbf{B}(r) = B_0 \left(\frac{r}{R_{\odot}} \right)^{-\delta} \mathbf{\hat{e}_r}\label{eq:single_B}
\end{eqnarray} 
where $B_0,N_0,\alpha, \delta$ are free parameters.  The constant $B_0$ can be of either polarity and reverses sign at the coronal current sheet.  The resulting expression for rotation measure RM is 
\begin{equation}\label{eq:RM_model_1}
\mathrm{RM} = \left[ \frac{2C_{FR} R_{\odot} N_0 B_0}{(\gamma -1) R_0^{\gamma-1}}  \right] \cos^{\gamma-1}\beta_c
\end{equation} 
where $C_{FR} \equiv e^3/2 \pi m_e^2 c^4, \gamma \equiv \alpha + \delta$, and $R_0$ is defined in Figure~\ref{fig:cartoon1} and given in solar radii.   The sign of the rotation measure depends on the polarity of $\mathbf{B}$ for $\beta < \beta_c$.  A positive sign obtains if $B_0 > 0$ for $\beta > \beta_c$, otherwise $RM < 0$.  Equation~\eqref{eq:RM_model_1} is Equation (4) from \cite{Ingleby:2007}.  The expression Equation~\eqref{eq:RM_model_1} is in cgs units.  For MKS units (the conventional units of rad/m$^2$), the number resulting from Equation~\eqref{eq:RM_model_1} should be multiplied by $10^4$.  

Equation~\eqref{eq:RM_model_1} shows that for $\beta_c = 0$, the rotation measure obtains its maximum possible value for a given $R_0$, and Equation~\eqref{eq:RM_model_1} becomes Equation (6) of \cite{Sakurai&Spangler:1994a}. This latter equation was used in \cite{Spangler:2005}.  Equation~\eqref{eq:RM_model_1} and the model which produced it account for time-variable Faraday rotation through the time dependence of  $\beta_c$ and $R_0$.  For the model of \cite{Sakurai&Spangler:1994a} we have $N_0 = 1.83 \times 10^6$ cm$^{-3}$, $\alpha=2.36$, $B_0 = 1.01$ G, and $\delta = 2$.  The predictions of Equation~\eqref{eq:RM_model_1} with the aforementioned constants and the two values for $\beta_c$ are shown in Table~\ref{T:Model_RM}.  The rotation measure has been calculated at times of 15:42, 19:35, and 23:19 UTC.  Discussion of the significance of the comparison of data and model is deferred to Section~\ref{Sec:Model_sum} below.


\subsubsection{Model 2 - Uniform Corona with Dual Power Laws in Density and Magnetic Field}\label{Sec:Model_2}

Within heliocentric distances of $<5 R_\odot$, the dependence of plasma density on heliocentric distance is more complex than a single power law relationship.  Similarly, the magnetic field is better represented by a dual power law that contains a dipole term ($\propto r^{-3}$) and an interplanetary magnetic field term ($\propto r^{-2}$).  Consequently, more accurate expressions for the heliocentric distance dependence of plasma density and magnetic field strength can be had with a compound power law for density, such as that used by \cite{Mancuso&Spangler:2000}, who adopted such expressions from \cite{Guhathakurta:1999} and \cite{Gibson:1999}, and a compound form of $B(r)$ from \cite{Patzold:1987}.  Expressions of this form for the case of dual power laws can be written
\begin{eqnarray}
n_e(r) = N_0 \left[ A \left(\frac{r}{R_{\odot}} \right)^{-\alpha_1} + B \left( \frac{r}{R_{\odot}} \right)^{-\alpha_2} \right]\label{eq:double_ne}\\
\mathbf{B}(r) = B_0 \left[ C \left(\frac{r}{R_{\odot}} \right)^{-\delta_1} + D \left( \frac{r}{R_{\odot}} \right)^{-\delta_2} \right] \mathbf{\hat{e}_r}\label{eq:double_B}
\end{eqnarray}
where Equations~\eqref{eq:double_ne} and \eqref{eq:double_B} have been written to resemble \eqref{eq:single_ne} and \eqref{eq:single_B} as closely as possible, and $A,B,C,D$ are dimensionless, adjustable constants.  

Equations~\eqref{eq:double_ne} and \eqref{eq:double_B} can be substituted into Equation~\eqref{eq:FR}, the variables $r$ and $s$ again converted to $R_0$ and $\beta$, and an expression obtained in terms of elementary angular integrals.  The result is
\begin{eqnarray}\label{eq:RM_model_2}
\mathrm{RM} = 2 C_{FR} R_{\odot} N_0 B_0  \sum^4_{i=1}\left(\frac{K_i}{(\gamma_i-1) R_0^{\gamma_i-1}} \right)\cos^{\gamma_i-1}\beta_c
\end{eqnarray}  
where $K_1 = AC, K_2 = AD, K_3 = BC, \mbox{ and } K_4 = BD$ and, correspondingly, $\gamma_1=\alpha_1+\delta_1,\gamma_2=\alpha_1+\delta_2,\gamma_3=\alpha_2+\delta_1, \mbox{ and }\gamma_4=\alpha_2+\delta_2$.  

Equation~\eqref{eq:RM_model_2} is of the same form as Equation~\eqref{eq:RM_model_1}, except it is polynomial in $R_0$ and $\cos \beta_c$, and produces a greater range of RM for the same range in $R_0$.  This model is defined by the adjustable constants $N_0,B_0,A,B,C, \mbox{ and }D$, as well as $\alpha_1,\alpha_2,\delta_1,\mbox{ and }\delta_2$.  We adopt the values used in \cite{Mancuso&Spangler:2000}, which were taken from \cite{Gibson:1999} and \cite{Patzold:1987}.  These values are $N_0 = 3.6 \times 10^5 \mbox{ cm}^{-3}, B_0 = 1.43 \mbox{ G}, A=101, B=1, C=4.17, \mbox{ and } D=1$.  The power law indices in density and magnetic field we adopt are $\alpha_1=4.3,\alpha_2=2,\delta_1=3,\mbox{ and }\delta_2=2$.

It should be emphasized that $N_0$ and $B_0$ are chosen as constants to accurately model the density and magnetic field (via Equations~\eqref{eq:double_ne} and \eqref{eq:double_B}) in the region of interest, $4.6 R_{\odot} \leq r \leq 10.0 R_{\odot}$.  Although $N_0$ and $B_0$ appear algebraically as the photospheric density and magnetic field in Equations~\eqref{eq:double_ne} and \eqref{eq:double_B}, we recognize that the true values at $r=1.0 R_{\odot}$ are much higher. The true density and magnetic field strength are steeper functions of $r$ than \eqref{eq:double_ne} and \eqref{eq:double_B} inside the region of space probed by the Faraday rotation measurements.  

The predictions of the model given by Equation~\eqref{eq:RM_model_2}, using the values of $R_0$ and $\beta_c$ are shown in Table~\ref{T:Model_RM}.  The values for Model 2 are of the same sign as Model 1, but with larger values of $|\text{RM}|$.  For August 17, the predicted RM exceeds the observed values by as much as an order of magnitude for the whole session.  As discussed in \cite{Spangler:2005}, this is partially due to higher values for the coronal magnetic field and plasma density, Equations~\eqref{eq:double_ne} and \eqref{eq:double_B}, than in Model 1.


\subsubsection{Model 3 - Corona with Streamers and Coronal Holes}\label{Sec:Model_3}

Models 1 and 2 assume that the  coronal density (and magnetic field strength) depend only on the heliocentric distance $r$.  While such a simplification may be acceptable at times of solar maximum, when the corona is approximately spherically symmetric, it is not generally justifiable because of the division of the corona into dense streamers and underdense coronal holes.  This is the situation most relevant to our observations.

\begin{figure}[htb!]
	\begin{center}
	\includegraphics[height=2.5in, trim = {25mm 35mm 23mm 35mm}, clip]{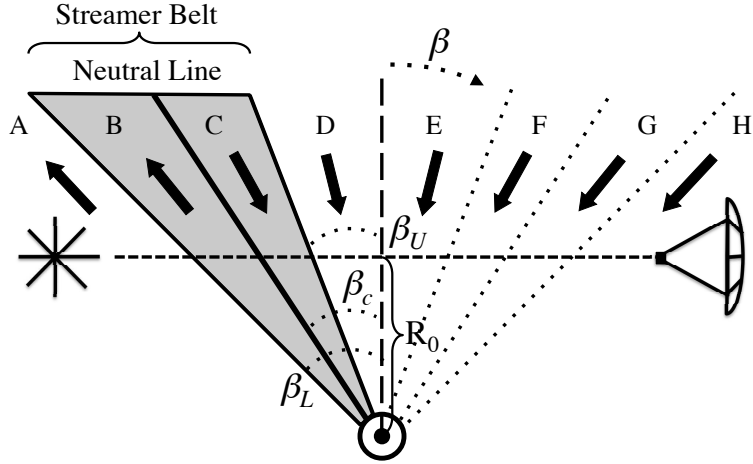}
	\caption{Geometry of the line of sight in coronal Model 3. Block arrows indicate the direction of the magnetic field.  The heavy lines define the sector boundaries for the high density coronal streamer belt (the shaded region, sectors B and C).  The angles $\beta_{L}$ and $\beta_{U}$ define the lower and upper bounds of the streamer belt.  The dashed line divides the line of sight in two halves of equal length and the dotted lines indicate symmetry lines.}
	\label{fig:cartoon2}
	\end{center}
\end{figure}

In considering the contribution of various parts of the line of sight to the rotation measure, one must pay attention not only to the angular position of coronal neutral line $\beta_c$, but also to the extent of the coronal streamer along the line of sight (again because it is much more dense than the surrounding plasma).  The geometry of the situation is illustrated in Figure~\ref{fig:cartoon2}.  In computing RM from Equation~\eqref{eq:FR}, the integral over $\beta$ is divided into 4 domains: $-\frac{\pi}{2} < \beta < \beta_{L}, \beta_{L} < \beta < \beta_c, \beta_c < \beta < \beta_{U}, \mbox{ and } \beta_{U} < \beta < \frac{\pi}{2}$.  The angle $\beta_c$ is defined as before, and $\beta_{L}$ and $\beta_{U}$ define the lower and upper limits of the streamer belt.  

In calculating the rotation measure for Model 3, we adopt the following model for the density and magnetic field.  For the density in the coronal hole, we have 
\begin{equation}
n_h(r) =  N_{0h} \left[ A_h \left(\frac{r}{R_{\odot}} \right)^{-\alpha_1} + B_h \left( \frac{r}{R_{\odot}} \right)^{-\alpha_2} \right]\label{eq:hole_ne}
\end{equation}
For the density in the higher density coronal streamer, we adopt 
\begin{equation}
n_s(r) =  N_{0s} \left[ A_s \left(\frac{r}{R_{\odot}} \right)^{-\alpha_1} + B_s \left( \frac{r}{R_{\odot}} \right)^{-\alpha_2} \right]\label{eq:streamer_ne}
\end{equation}
We assume, for reasons of simplicity in algebraic bookkeeping, that the indices $\alpha_1$ and $\alpha_2$ are the same in the streamers and coronal holes.  The differences in the plasma density of the streamers and holes are then controlled by the parameters $N_{0h}, N_{0s}, A_h, B_h, A_s, \mbox{ and } B_s$. In the calculations to be presented below, we used the same model for $n_s(r)$ as in Section~\ref{Sec:Model_2}.  For the parameters $N_{0h},A_h,\mbox{ and }B_h$ we used values of $N_{0h}=1.01 \times 10^{5}\mbox{ cm}^{-3}, A_h=25.85, \mbox{ and } B_h=1$.  Quantities which describe coronal hole-to-streamer contrast are $N_{0h}/N_{0s}=0.305\mbox{ and }A_h/A_s=0.256\mbox{ and }B_h/B_s=1$.  These values are chosen to closely correspond to the coronal hole model of \cite{Mancuso&Spangler:2000}.

The magnetic field is assumed to be the same in the streamer and coronal holes and, as before, assumed to be entirely radial and to change sign at the current sheet.  The algebraic form is assumed to be of the same form as Equation~\eqref{eq:double_B}.  Equations~\eqref{eq:hole_ne},~\eqref{eq:streamer_ne}, and~\eqref{eq:double_B} are then substituted into~\eqref{eq:FR} and the integral over $\beta$ carried out over the limits identified above.  The resultant expression is the same as Equation~\eqref{eq:RM_model_2} 
\begin{eqnarray}\label{eq:RM_model_3}
\mathrm{RM} = 2 C_{FR} R_{\odot} N_{0s} B_0  \sum^4_{i=1}\left(\frac{K_i}{(\gamma_i-1) R_0^{\gamma_i-1}} \right)\left[\cos^{\gamma_i-1}\beta_c-W_i(\beta_L,\beta_U)\right]
\end{eqnarray}  
except there is a correction factor $W_i(\beta_L,\beta_U)$ due to the finite extent of the coronal streamer given by
\begin{eqnarray}\label{eq:RM_model_3_W}
W_i(\beta_L,\beta_U) = \frac{1}{2}\left(1-\frac{N_{0h}}{N_{0s}}\xi_i\right)\left\{\cos^{\gamma_i-1}\beta_L+\cos^{\gamma_i-1}\beta_U\right\}
\end{eqnarray}
Similar to Equation~\eqref{eq:RM_model_2}, where $K_1 = A_sC, K_2 = A_sD, K_3 = B_sC, \mbox{ and } K_4 = B_sD$ and, as in Equation~\eqref{eq:RM_model_2}, $\gamma_1=\alpha_1+\delta_1,\gamma_2=\alpha_1+\delta_2,\gamma_3=\alpha_2+\delta_1, \mbox{ and }\gamma_4=\alpha_2+\delta_2$.  The new parameter here, $\xi_i$, is given by $\xi_1=\xi_2=A_h/A_s$ and $\xi_3=\xi_4=B_h/B_s$.  In addition to the parameters already identified for Models 1 and 2, the expression for the rotation measure also depends on the angles $\beta_{L} \mbox{ and } \beta_{U}$ and parameters $A_h N_{0h}/A_s N_{0s}\mbox{ and } B_h N_{0h}/B_s N_{0s}$ which describe the density contrast of the coronal streamer and holes.

The angular limits of the streamer, $\beta_{U}$ and $\beta_{L}$ were determined by assuming the half-thickness of the streamer, $\theta$, (perpendicular to the neutral line) is $12\degr$ \citep{Bird:1996,Mancuso&Spangler:2000}.  We then have $\beta_{L} = \beta_c - \theta, \beta_{U} = \beta_c + \theta$.  We have not attempted to account for the varying thickness of the streamer belt with position along the neutral line, as was done by \cite{Mancuso&Spangler:2000} on the basis of contemporanous data.  They found that this improved the degree of agreement between the model and observed RMs.  Unlike \cite{Mancuso&Spangler:2000}, we are not trying to determine the best fit parameters for this model, we are only interested in trying to understand why the measured RM was lower than we anticipated.

The predictions from Model 3 are shown in Table~\ref{T:Model_RM}.  Like Models 1 and 2, Model 3 reproduces the sign of the observed RM at the beginning of the session.  The key difference is that Model 3 provides the best estimate of the magnitude of the RM observed at the beginning of the session.  This reduction in magnitude compared to Models 1 and 2 can be understood by reference to Figure~\ref{fig:cartoon2}.  From symmetry, the contributions to the rotation measure from sectors D and E cancel.  Sectors C and F, which cancel in the case of a uniform corona (the case of Models 1 and 2, illustrated in Figure 1), do not in the present case because the density in sector C is higher than in F.  Sector C makes a positive contribution (for the polarity shown in the figure) to the RM which is larger in absolute magnitude than that of F.  Sectors A, B, G, and H all make a negative contribution to the rotation measure.  The magnitude of the total (observed) rotation measure is then reduced because of the positive contribution by sector C in the high density coronal streamer belt.

It is interesting to note that Model 3 allows for a positive RM for the geometry of this line of sight (albeit for a different set of parameters than those we have chosen in Table~\ref{T:Model_RM}), whereas Models 1 and 2 will only yield negative RM.  This results because the sign of the total (observed) rotation measure depends on whether the RM calculation of $|C| > A+B+F+G+H$.  Sector B is the only one of the negative RM sectors in the denser current sheet.  It is located at larger values of $\beta$ than the positive sector C, so its absolute magnitude will always be less than that of C.  If the current sheet is modeled as much denser than the coronal hole, as was the case in \cite{Mancuso&Spangler:2000} and the calculation above, the net rotation measure due to sectors B+C will be positive.  Whether the total RM is positive or negative is determined by its functional dependence on $\beta$, which in turn is determined by the radial gradients in plasma density and magnetic field (the power law indices $\alpha$ and  $\delta$ which determine $\gamma$), and by the density contrast between the streamer and coronal hole.  


\subsection{Summary of Analytic Coronal Models}\label{Sec:Model_sum}

The results of the analytic coronal Models 1-3 (Sections~\ref{Sec:Model_1} - \ref{Sec:Model_3}) are shown on the right hand side of Table~\ref{T:Model_RM} and may be compared to the data as follows. 
\begin{enumerate}
\item Models 1 and 2, which assume that the same power laws for density and magnitude of the magnetic field apply at all points along the line of sight, correctly give the sign of the RM at the beginning of the observing session.  However, they are not able to reproduce the observed magnitude of the RM; both models overestimate the magnitude.  These two models do not give the right sign for $d/dt(\mathrm{RM}) = \dot{\mathrm{RM}}$ and they both underestimate the magnitude of the variation in the observed RM.

\begin{table}[htdp]
	\begin{center}
	\caption{Model rotation measures}\label{T:Model_RM}
	\medskip
	\begin{threeparttable}
		\begin{tabular}{ccc|cccc}
		\hline\hline
		Time (UTC)	&	$R_0$ 	&	RM\tablenotemark{a}		&	$\beta_c$\tablenotemark{b}	&	Model 1\tablenotemark{c}		&	Model 2\tablenotemark{d}		&	Model 3\tablenotemark{d}\\
					&			& [rad/m$^2$]	&							&	[rad/m$^2$]	&	[rad/m$^2$]	&	[rad/m$^2$]\\
		\hline	
		15:42		&	5.00		&	$-21.2\pm1.66$	&	$- 33.0\degr$				&	$- 50.0$		&	$-89.5$		&	$- 12.8$\\
					&			&	$-12.7\pm1.48$	&	$- 42.3\degr$				&	$- 32.8$		&	$-51.5$		&	$- 2.7$\\
		19:35		&	4.76		&	$9.83\pm1.63$		&	$- 33.2\degr$				&	$- 58.5$		&	$-110.7$		&	$- 14.5$\\
					&			&	$9.09\pm1.46$		&	$- 41.9\degr$				&	$- 39.5$		&	$-65.5$		&	$- 2.8$\\
		23:19		&	4.58		&	$6.07\pm1.48$		&	$- 33.5\degr$				&	$- 65.8$		&	$-130.1$		&	$- 15.7$\\
					&			&	$4.31\pm1.38$		&	$- 41.8\degr$				&	$- 45.2$		&	$-78.3$		&	$- 2.6$\\
		\hline
		\end{tabular}
	\medskip
		\begin{tablenotes}
		\small
		\item[a] The upper and lower values correspond to the northern and southern hotspots, respectively.
		\item[b] The upper and lower values were calculated using Carrington rotation 2113 and 2114, respectively.
		\item[c] The model parameters are taken from \cite{Sakurai&Spangler:1994a}.
		\item[d] The model parameters are taken from \cite{Mancuso&Spangler:2000}.
		\end{tablenotes}	
	\end{threeparttable}
	\end{center}
\end{table}%

\item Model 3, which attempts to more accurately account for the division of the corona into the coronal holes and streamers, approximately reproduces the magnitude of the observed rotation measure during the beginning of the observations for $\beta_c\approx-33\degr$.  For $\beta_c\approx-42\degr$, though, Model 3 gives an almost constant value of $-2.7$ rad/m$^2$.  Like Models 1 and 2, Model 3 can not reproduce the observed variations in the RM.
\item The magnetic field model given in Equation~\eqref{eq:single_B}, which gives a coronal field of 28 mG at a fiducial heliocentric distance of $6 R_{\odot}$, or Equation~\eqref{eq:double_B}, which gives a value of 67 mG, are equally compatible with the data near the beginning of the observing session.  Either of these is compatible with the range of values presented in \cite{Dulk&McLean:1978}, and the more recent summaries of \cite{You:2012} and \cite{Mancuso&Garzelli:2013}.
\item The results of Figure~\ref{fig:RM_Time_Series} and these modeling cases suggest that we can not make a precision measurement of the coronal magnetic field strength using our observations because these models can not reproduce the entire rotation measure time series.  Consequently, we have not tried to determine best fit model parameters for the magnetic field models in Equations~\eqref{eq:single_B} and~\eqref{eq:double_B}.
\item  Our results emphasize the importance of studying RM($t$) time series when remotely sensing the corona, rather than maximum values of RM or use of the RM at one time in the session, as done in \cite{Spangler:2005}.  For the conditions present in the corona during our observations, we measured a much smaller RM than indicated by Models 1 and 2, but of the order indicated by Model 3.  This suggests that, to make more precise measurements of the magnetic field in the corona, the observer must carefully consider the conditions that will be present on or near the day of observation.  In particular, future observations should be dynamically scheduled such that the lines of sight are occulted by the streamer belt in such a way as to minimize $\beta_c$ and, therefore, maximize the observed Faraday rotation.
\end{enumerate}


\section{Implications for Coronal Heating}\label{sec:discussion}

\subsection{Differential Rotation Measures and Constraints on Joule Heating}\label{sec:discuss_Diff_Far}

One of the most striking features of Figure~\ref{fig:RM_Time_Series} is the smooth variation of RM as a function of time and the similarity of this time series for both components.  However, there are several periods during which the Faraday rotation observed for the southern hotspot is not equal to that of the northern hotspot.  This highlights one of the advantages of using extended extragalactic radio sources for Faraday rotation measurements: we can measure differential Faraday rotation, $\Delta\mathrm{RM}$.  The observational hallmark of differential Faraday rotation is a difference in the Faraday rotation to two or more components in a radio source, a clear example of which is shown in Figure~\ref{fig:DELTA_X_Scan_10}.  

\begin{figure}[htb!]
	\begin{center}
	\includegraphics[width=4.25in]{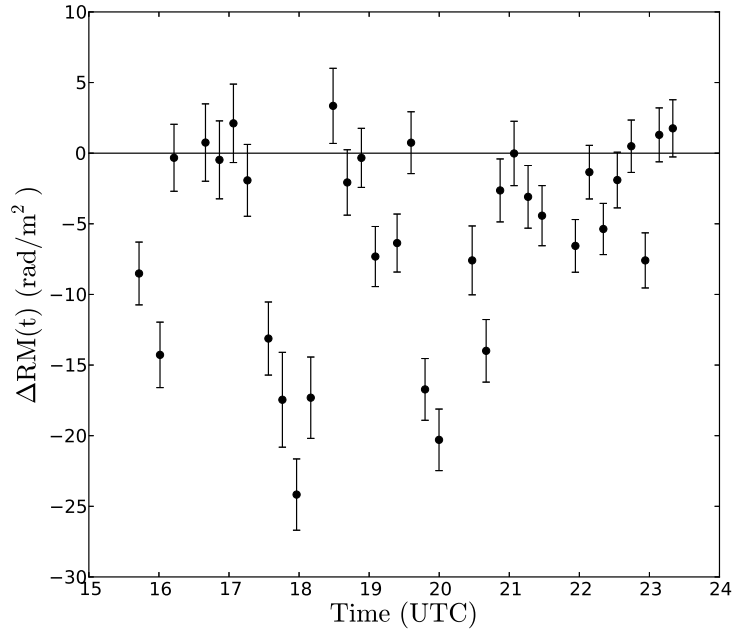}
	\caption{Differential Faraday rotation data as a function of time.  The measurement shows the difference in the rotation measures for the northern and southern hotspots of 3C228, which are separated by 46$\arcsec$ (33,000 km in the corona).}
	 \label{fig:Diff_RM_Time_Series}
	\end{center}
\end{figure}

The data in Figure~\ref{fig:RM_Time_Series} have been used to calculate the differential Faraday rotation time series, $\Delta\mathrm{RM}(t)$, by subtracting the RM($t$) time series for the southern hotspot from RM($t$) for the northern hotspot.  These data are shown in Figure~\ref{fig:Diff_RM_Time_Series}.
As with Figure~\ref{fig:RM_Time_Series}, the values presented here use data from all observing frequencies; the errors are calculated from a propagation of known radiometer noise errors and may (and probably do) underestimate the true errors.  For most of the scans, the $\Delta\mathrm{RM}$ measurement is consistent with zero; however, there are three periods of prominent $\Delta\mathrm{RM}$ in Figure~\ref{fig:Diff_RM_Time_Series}: 15:37 - 16:05, 17:28 - 18:14, and 19:42 - 20:44 UTC.  The differential rotation measure for the scan with the maximum $\Delta\mathrm{RM}$ during each period is given in Table~\ref{T:currents}.  
\begin{table}[htdp]
	\begin{center}
	\caption{Approximate Coronal Currents}\label{T:currents}
	\medskip
	\begin{threeparttable}
		\begin{tabular}{llcc}
		\hline\hline
		Time (UTC)	&	$R_0$ 	&	$\Delta\rm{RM(t)}$ [rad/m$^2$]		&	$I_{obs}$\tablenotemark{a}	[GA]\\
		\hline	
		16:00			&	4.98		&	$- 14.3 \pm 2.3$						&	$2.6\pm0.4$\\
		17:57			&	4.86		&	$- 24.2 \pm 2.5$						&	$4.1\pm0.4$\\
		19:59			&	4.74		&	$- 20.3 \pm 2.2$						&	$3.3\pm0.4$\\
		\hline
		\end{tabular}
	\medskip
		\begin{tablenotes}
		\small
		\item[a] Calculated with $\alpha=1/2$.
		\end{tablenotes}	
	\end{threeparttable}
	\end{center}
\end{table}%
In all three cases, the differential Faraday rotation events persisted for at least two scans ($\sim 30$ minutes).  The differential Faraday rotation demonstrated in Figure~\ref{fig:DELTA_X_Scan_10} corresponds to the scan at 17:57 UTC in Table~\ref{T:currents} and represents the maximum differential rotation measure obtained during these observations.  It is interesting, though perhaps coincidental, that these instances of differential Faraday rotation occur at times when there are changes in the slope of RM($t$).  Examination of the $\Delta\mathrm{RM}(t)$ from the second set of maps of 3C228 which averaged together three or four scans, again, did not reveal any qualitatively or quantitatively new results.

Provided that this technique is valid for astrophysical plasmas and not merely for measuring laboratory plasmas, our data on differential Faraday rotation yield interesting information on electrical currents in the corona.  As discussed in Section~\ref{sec:Heating_Models}, \cite{Spangler:2007,Spangler:2009} demonstrated that the differential rotation measure can be used to estimate the magnitude of the coronal currents flowing between the lines of sight.  As shown in Figure 1 of that paper, an Amperian loop can be set up along the two lines of sight that allows computation of the current, $I$, via Ampere's law.  Crucial in this derivation is the assumption that the measured $\Delta\mathrm{RM}$ is dominated by a region in which the plasma density is relatively uniform; for detailed discussion of the validity of this approximation, see \cite{Spangler:2007}.  Using the single power law of Equation~\eqref{eq:single_ne} to estimate this uniform density, gives the following approximation for the current magnitude (in SI units)
\begin{eqnarray}\label{eq:current_SI}
I_{obs} = 3.3\times10^6\left(\Delta\rm{RM}\right)R_0^{2.5}\hspace{3mm}\rm{A}
\end{eqnarray}
which is Equation (7)\footnote{\cite{Spangler:2007} also includes a plasma density scaling parameter, $\alpha\in[0,1]$, that results from the assumption of a relatively uniform density.  In Equation~\eqref{eq:current_SI} and Table~\ref{T:currents}, we have used the same value used in \cite{Spangler:2007}, $\alpha=1/2$.} in \cite{Spangler:2007}.

We can use the data in Figure~\ref{fig:Diff_RM_Time_Series} in Equation~\eqref{eq:current_SI} to obtain estimates for the electrical current. The results for the three $\Delta\mathrm{RM}$ events are given in Table~\ref{T:currents}.  Table~\ref{T:currents} shows that detectable differential Faraday rotation requires currents of $10^9-10^{10}$ A.  These currents are of the same order as those reported by \cite{Spangler:2007}.  Assuming the detected differential Faraday rotation is indeed produced by coronal currents, then we reiterate the conclusion reported in \cite{Spangler:2007,Spangler:2009}: the currents we have detected are either irrelevant for Joule heating of the corona or the true resistivity in this nearly collisionless plasma is $5-6$ decades greater than the Spitzer resistivity.


\subsection{Rotation Measure Fluctuations and Constraints on Wave Heating}

\begin{figure}[htb!]
	\begin{center}
	\includegraphics[width=4.25in]{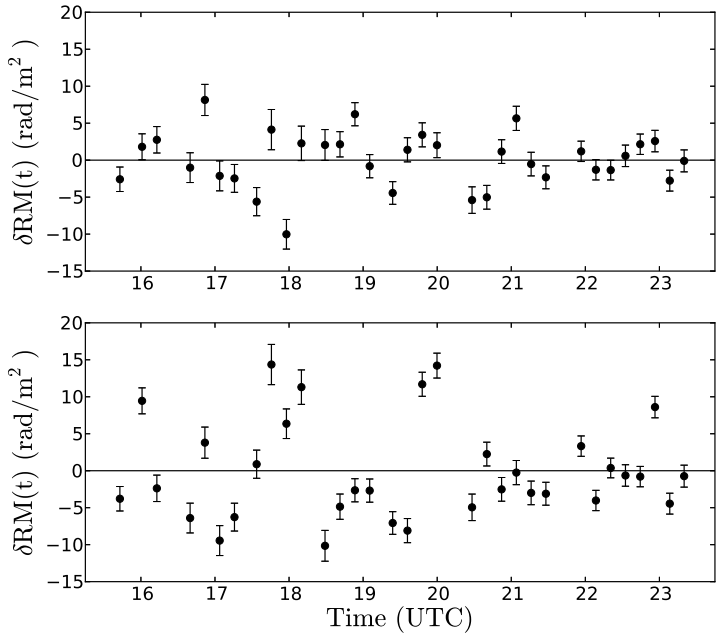}
	\caption{Rotation measure fluctuations for 3C228 on August 17, 2011, for the northern hotspot ({\it top}) and the southern hotspot ({\it bottom}).  Each point represents the RM for a given scan ($\sim9$ minutes duration) minus the smoothly varying fit shown in Figure~\ref{fig:RM_Time_Series}.}
	 \label{fig:RM_fluctuations}
	\end{center}
\end{figure}

As discussed in Section~\ref{sec:Heating_Models}, \cite{Hollweg:1982} demonstrated that fluctuations in the rotation measure time series, $\delta\mathrm{RM}(t)$, provide information (primarily) on fluctuations in the magnetic field.  We calculated $\delta\mathrm{RM}(t)$ directly by taking the data in Figure~\ref{fig:RM_Time_Series} and subtracting the best-fit quintic polynomial.  The resulting data are shown in Figure~\ref{fig:RM_fluctuations}.
One can determine the extent to which these residual fluctuations are caused by true coronal RM variations in the signal versus random fluctuations due to noise by examining the auto-correlation and cross-correlation functions for these two signals.  If the rotation measure fluctuations are caused by coronal fluctuations (either in plasma density or in the magnetic field lines, e.g., Alfv{\' e}n waves), then each time step in the $\delta\mathrm{RM}(t)$ series should be partially correlated with previous and future time steps.  If, however, the fluctuations are due to noise, we expect no such correlation between time steps.

\begin{figure}[htb!]
	\centering
	\subfigure[]{
		\includegraphics[width=2.5in]{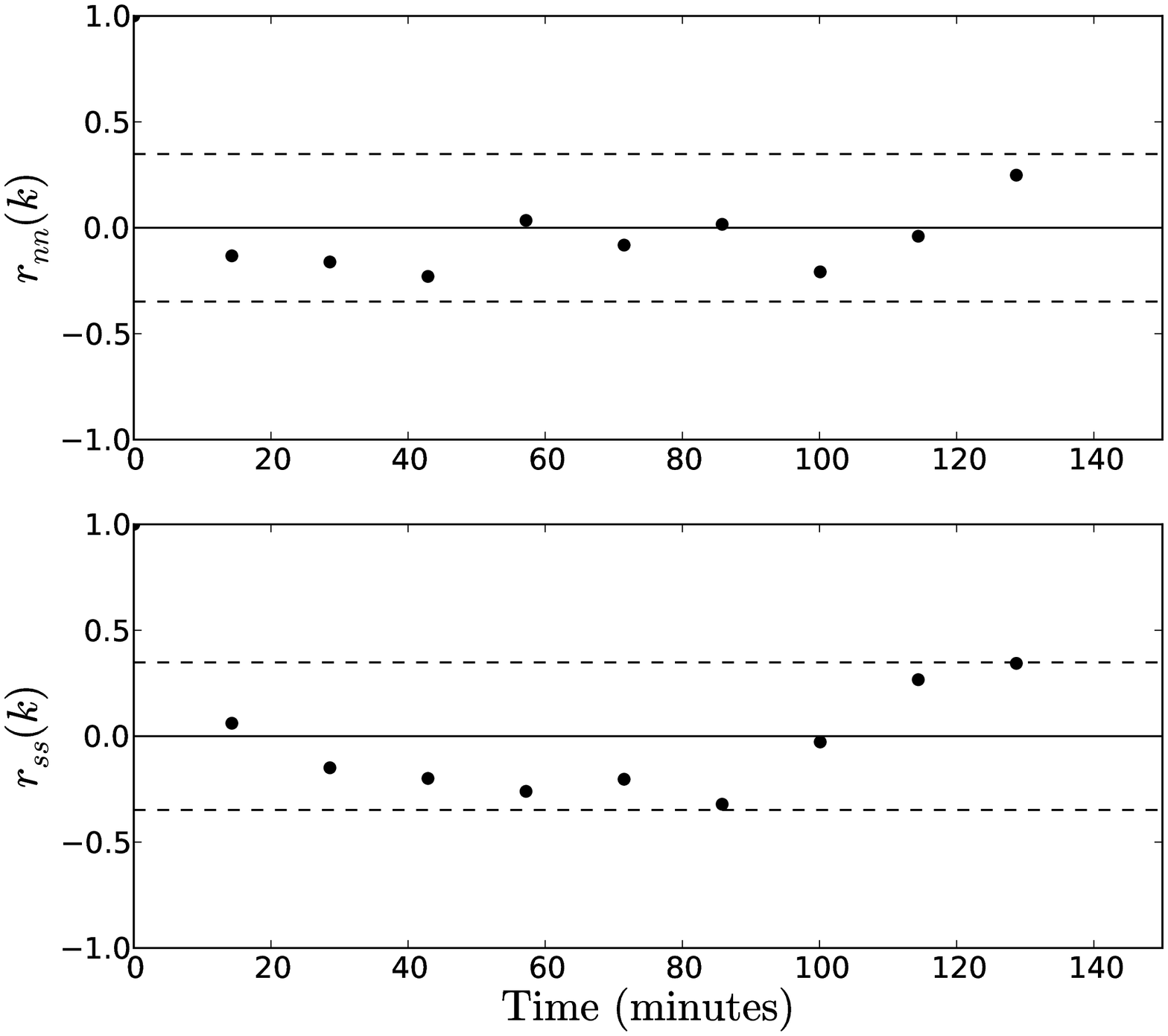}
		\label{fig:RM_fluctuations_autocorr}
          }
	\hspace{.1in}
	\subfigure[]{
		\includegraphics[width=2.5in]{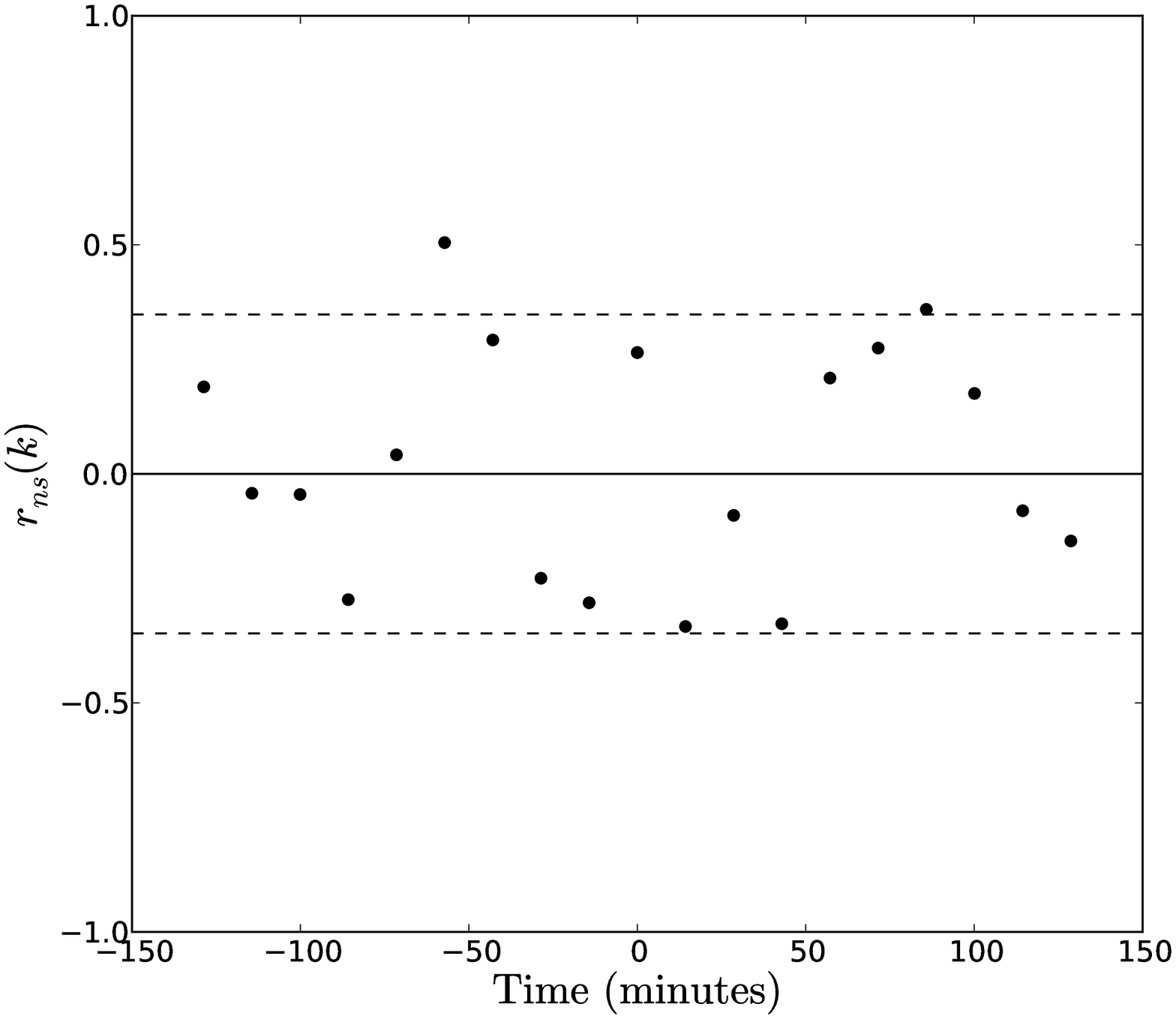}
		\label{fig:RM_fluctuations_crosscorr}
	}
	\caption[]{(a) The auto-correlation function of the rotation measure fluctuations for the northern hotspot ({\it top}) and the southern hotspot ({\it bottom}).  (b) The cross-correlation function of the northern and southern hotspots.  The dashed lines represent the 95\% confidence limits for a random series.
  }
\label{fig:auto_and_cross}
\end{figure}

Figure~\ref{fig:RM_fluctuations_autocorr} and~\ref{fig:RM_fluctuations_crosscorr} show the auto-correlation and cross-correlation functions for the northern and southern hotspots out to lag 9.  The lag $k$ values have been converted to time lag values (lag 1 corresponds to $\sim14$ minutes).  The dashed lines in Figure~\ref{fig:auto_and_cross} represent the 95\% confidence band.  For $N$ independent, identically distributed random variables, the lagged values are uncorrelated and the expectation value and variance of the auto-correlation function are $0$ and $1/N$, respectively.  The 95\% confidence levels are then $\pm2/\sqrt{N}$.  
For our observations, $N=33$.  Because the values for the auto- and cross-correlation functions generally lie within this band, the data are consistent with random noise.  The lag $-4$ ($-57.2$ minutes) value for the cross-correlation function does lay outside of this band in Figure~\ref{fig:RM_fluctuations_crosscorr}; however, by definition, we expect $\sim1$ value out of twenty to be outside the $95\%$ confidence band.

Because we see neither an auto-correlated nor a cross-correlated signal in the rotation measure fluctuations, we can use the data to place an upper limit on the magnitude of observable RM fluctuations due to magnetic field fluctuations, $\delta B$, in the corona.  The variance of the data, $\sigma_{RM}^2$, in Figure~\ref{fig:RM_fluctuations} can be modeled as the sum of the variance due to known radiometer noise, $\sigma_{noise}^2$, and the variance due to coronal fluctuations, $\sigma_{C}^2$:
\begin{eqnarray}\label{eq:variance_fluc}
\sigma_{RM}^2 = \sigma_{noise}^2+\sigma_{C}^2
\end{eqnarray}
These values are given in Table~\ref{T:fluctuations}.  
\begin{table}[htdp]
	\begin{center}
	\caption{Estimations for Rotation Measure Fluctuations}\label{T:fluctuations}
	\smallskip
	\begin{threeparttable}
		\begin{tabular}{lcccc}
		\hline\hline
		Source				&	$\sigma_{RM}$		&	$\sigma_{noise}$	&	$\sigma_{C}$	 	& 	$\delta\chi$\tablenotemark{a}\\
							&	[rad/m$^2$]		&	[rad/m$^2$]		&	[rad/m$^2$]			&	[degrees]		\\
		\hline	
		Northern Hotspot	&	3.7 (3.3)\tablenotemark{b}	&	1.7 (1.7)		&	3.3 (2.8)		&	3.2 (2.8)	\\
		Southern Hotspot	&	6.6 (6.2)	&	1.5 (1.5)		&	6.4 (6.0)		&	6.3 (5.9)	\\		
		Sakurai and Spangler (1994a)\tablenotemark{c} & 1.6 & ... & ... & 1.6 \\
		Mancuso and Spangler (1999)\tablenotemark{d} & 0.40 & ... & ... & 0.39 \\
		\hline
		\end{tabular}
	\medskip
		\begin{tablenotes}
		\small
		\item[a] Estimation using a fiducial frequency of 2.3 GHz.
		\item[b] All values in parentheses are calculated by removing the residual with the largest magnitude.
		\item[c] Measurements were taken at an average impact parameter of $R_0\approx9.0R_\odot$.
		\item[d] Measurements were taken at an average impact parameter of $R_0\approx8.6R_\odot$.
		\end{tablenotes}	
	\end{threeparttable}
	\end{center}
\end{table}
The variance due to coronal fluctuations, $\sigma_{C}^2$, can then be used to estimate the Faraday rotation fluctuations in degrees, $\delta\chi$.  To compare our results with those of \cite{Hollweg:1982} and \cite{Andreev:1997a}, we chose a fiducial frequency of 2.3 GHz, giving $3\fdg2$ and $6\fdg3$ for the Northern and Southern hotspots respectively.  We also removed the RM fluctuation residuals with the largest $|\delta\mathrm{RM}|$ (scan 10 at 17:57 UTC for the northern hotspot, corresponding to the RM value demonstrated by the solid line in Figure~\ref{fig:DELTA_X_Scan_10}, and scan 9 at 17:45 UTC for the southern hotspot) and recalculated the variances.  These values appear in parentheses in Table~\ref{T:fluctuations}.  Removing the largest outlier in each $\delta\mathrm{RM}(t)$ does not have much of an effect; $\sigma_{noise}$ does not change, but $\sigma_{RM}$ decreases slightly and, consequently, $\sigma_{C}$ decreases by $\sim15\%$ and $\sim6\%$ for the northern and southern hotspots, respectively.  This calculation was also repeated removing the two largest residuals with similar results.  Thus, we conclude that the calculation is not dominated by one or two outliers.

For comparison, the upper limit from \cite{Sakurai&Spangler:1994a} and the possible detection reported in \cite{Mancuso&Spangler:1999} are included in Table~\ref{T:fluctuations}.  The values reported in \cite{Sakurai&Spangler:1994a} and \cite{Mancuso&Spangler:1999} are smaller in magnitude than our present results, but their measurements were performed at greater heliocentric distances.

\begin{figure}[htb!]
	\begin{center}
	\includegraphics[width=4.25in]{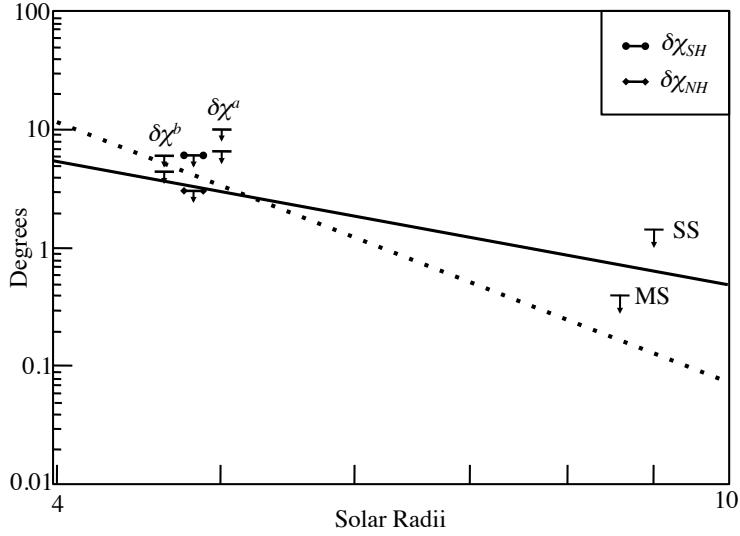}
	\caption{Faraday rotation fluctuations, $\delta\chi$, as a function of the impact parameter, $R_0$, in solar radii.  $\delta\chi$ is expressed in terms of the standard deviation of the position angle at 2.3 GHz.  $\delta\chi_{NH}$ and $\delta\chi_{SH}$ ($R_0\sim4.8$) are the upper limits obtained by removing the polynomial fits for the northern and southern hotspots, respectively, from our present study.  $\delta\chi^a$ ($R_0\sim5.0$) and $\delta\chi^b$ ($R_0\sim4.6$) are the upper limits obtained by removing the linear fits for the first and last five hours, respectively.  For both sets $\delta\chi^a$, and $\delta\chi^b$, the upper and lower bars are for data from the southern and northern hotspots, respectively.  The values for $R_0$ selected for $\delta\chi^a$, $\delta\chi^b$, $\delta\chi_{NH}$ and $\delta\chi_{SH}$ are a plotting convention only.  ``SS'' is an upper limit from \cite{Sakurai&Spangler:1994a} and ``MS'' is a possible detection taken as an upper limit from \cite{Mancuso&Spangler:1999}.  The solid line is a power law fit (exponent $-2.56$) to the observed values from \cite{Hollweg:1982} and the dotted line is a single power law fit (exponent $-5.50$) to observations adapted from \cite{Andreev:1997a}.  This is an updated version of Figure 9 from \cite{Mancuso&Spangler:1999}.}
	 \label{fig:mag_fluctuations_compare}
	\end{center}
\end{figure}

Figure~\ref{fig:mag_fluctuations_compare} is a reproduction of Figure 9 in \cite{Mancuso&Spangler:1999} and compares our present results to the observational results of \cite{Hollweg:1982}, \cite{Sakurai&Spangler:1994a}, \cite{Andreev:1997a}, and \cite{Mancuso&Spangler:1999}.  
The upper limits for Faraday rotation fluctuations for the northern and southern hotspots have been plotted as $\delta\chi_{NH}$ and $\delta\chi_{SH}$, respectively, at a fiducial impact parameter of $R_0\sim4.8$.  The upper limit from \cite{Sakurai&Spangler:1994a} is labeled ``SS'' and the possible detection reported in \cite{Mancuso&Spangler:1999} is given here as an upper limit ``MS.''  Also on this plot is a power law fit to the observed values from \cite{Hollweg:1982} (solid line) and a single power law fit adapted from the double power law fits to four sets of observations in \cite{Andreev:1997a} (dotted line).  

In comparing our values to \cite{Hollweg:1982} and \cite{Andreev:1997a}, it is important to note the different methods they used to calculate the Faraday rotation fluctuations.  The single power law fit (with exponent $-2.56$) from \cite{Hollweg:1982} was determined by removing a linear trend over a $\sim5$ hour tracking pass, whereas our values of $\delta\chi_{NH}$ and $\delta\chi_{SH}$ were calculated by removing a quintic polynomial over the eight hour session.  The original double power law fits in \cite{Andreev:1997a} were determined by integration over one decade (fluctuation periods: 1.5 - 15 minutes) of the temporal Faraday rotation fluctuation spectrum.  Further, the values reported in \cite{Hollweg:1982} were uncorrected for receiver noise because it was assumed to be negligible compared to the $100\%$ linearly polarized signal.  \cite{Andreev:1997a} corrected $\delta\chi$ for receiver noise at the specific ground stations for observation.  Our method is most similar to \cite{Hollweg:1982} and, to better compare our data with that of \cite{Hollweg:1982}, we also calculated additional estimates of the Faraday rotation fluctuations for the two hotspots by first removing a linear fit to (a) the first five hours $\delta\chi^a$ and (b) the last five hours $\delta\chi^b$.  In Figure~\ref{fig:mag_fluctuations_compare}, $\delta\chi^a$ and $\delta\chi^b$ are plotted at fiducial impact parameters of $R_0\sim5.0$ and $R_0\sim4.6$, respectively.  We chose these two periods because it is during the first five hours that the line of sight transits a bright coronal ray before entering and remaining in a coronal hole after $\sim$ 20:00 UTC.  The values for $R_0$ at which $\delta\chi^a$, $\delta\chi^b$, $\delta\chi_{NH}$, and $\delta\chi_{SH}$ have been plotted in Figure~\ref{fig:mag_fluctuations_compare} are a plotting convention only.

The Faraday rotation fluctuations calculated for the northern component using this method are $\delta\chi^a=6\fdg7$ and $\delta\chi^b=4\fdg4$; for the southern hotspot, they are $\delta\chi^a=10\fdg0$ and $\delta\chi^b=6\fdg2$.  We expect $\delta\chi^a$ and $\delta\chi^b$ to be greater than $\delta\chi_{NH}$ and $\delta\chi_{SH}$ for the northern and southern hotspots, respectively.  The RMS of residuals from a quintic fit to a time series is smaller than the RMS of residuals from a linear fit.  Our values of $\delta\chi_{NH}$ and $\delta\chi_{SH}$ are therefore biased low because the quintic polynomial does a better job accounting for the slow time variations in Figure~\ref{fig:RM_Time_Series}; however, we do not believe these slow time variations are coronal Alfv{\' e}n waves themselves because the period is extremely long.  

The southern hotspot, which has a stronger polarized intensity and, consequently, a smaller error associated with radiometer noise, yields $\delta\chi_{SH}\approx6\fdg3$ at an average impact parameter of $R_0 = 4.8R_\odot$ for the session.  This upper limit is consistent with the values expected from both \cite{Hollweg:1982} and \cite{Andreev:1997a} at this impact parameter.  Similarly, $\delta\chi^a$ and $\delta\chi^b$ for the southern hotspot (the upper bars in Figure~\ref{fig:mag_fluctuations_compare}) are consistent with \cite{Hollweg:1982} and \cite{Andreev:1997a}.

The northern hotspot gives $\delta\chi_{NH}\approx3\fdg2$ at an average impact parameter of $R_0 = 4.8R_\odot$.  While this upper limit is below the two fits, there is considerable variation in the data used to produce these fits.  The limit $\delta\chi^a$ (the lower bar in Figure~\ref{fig:mag_fluctuations_compare}) for the northern hotspot is above the two fits, while $\delta\chi^b$ (again, the lower bar in Figure~\ref{fig:mag_fluctuations_compare}) is above \cite{Hollweg:1982} and below \cite{Andreev:1997a}.  Consequently, the limits obtained from the northern component are not inconsistent with the observations of \cite{Hollweg:1982} and \cite{Andreev:1997a} at impact parameters of $4.6-5.0R_\odot$.

These results should also be compared with the results of \cite{Hollweg:2010}.  They applied the MHD wave-turbulence model of \cite{Cranmer:2007} to conditions within the corona during the {\it Helios} observations and, assuming a line of sight through an equatorial streamer, calculated the predicted Faraday rotation fluctuations.  Similar to the power law fit to data from \cite{Hollweg:1982}, their model \cite[``streamer'' in Figure 3 of][]{Hollweg:2010} was within the upper limit imposed by \cite{Sakurai&Spangler:1994a} but not within the limit given by \cite{Mancuso&Spangler:1999}.  Their ``streamer'' model, however, agrees well with the limits imposed by our observations.

It is nonetheless disappointing that several VLA observations undertaken since 1990 have not yielded clear detections of RM fluctuations on time scales of tens of minutes to approximately an hour.  The observations of \cite{Hollweg:1982} and \cite{Andreev:1997a} show a range of $\delta\mathrm{RM}$ values at a given impact parameter or solar elongation.  The RMS $\delta\chi$ varies by an order of magnitude at $5 R_\odot$ over a tracking pass in \cite{Hollweg:1982} and the $\delta\chi$ corresponding to ingress and egress during the 1983 occultation of {\it Helios} reported in \cite{Andreev:1997a} were also very different.  Upper limits or weak detections reported in \cite{Sakurai&Spangler:1994a}, \cite{Mancuso&Spangler:1999}, and the present paper are not inconsistent with the mean of these fluctuations, but a value of $\delta\mathrm{RM}$ substantially above the mean would have been clearly detected.

One advantage of the {\it Helios} observations was that the spacecraft were exactly in the ecliptic plane and therefore sampled coronal regions at low heliographic latitudes.  It is at these lower latitudes that the line of sight is more likely to transit an active region and, possibly, detect higher intensity Faraday rotation fluctuations.  Our present observations sampled heliographic latitudes ranging from $59\fdg4$ to $71\fdg0$, with the ray path transiting a coronal ray during the first four hours of observation before entering a coronal hole for the remaining four hours.  In view of the importance of the properties of coronal Alfv{\' e}n waves to heliospheric physics, we feel additional VLA searches for RM variations associated with these waves would be worthwhile.


\section{Summary and Conclusions}\label{sec:Summary}

\begin{enumerate}
\item Very Large Array (VLA) polarimetric observations of the extended, polarized radio source 3C228 were made for eight hours on August 17, 2011, when the line of sight was occulted by the solar corona, passing heliocentric distances (our parameter $R_0$) as small as $4.6 R_{\odot}$.  During this session, 33 scans of $\sim9$ minutes duration were made.  These observations yielded information on the rotation measure (RM) to the northern and southern hotspots, which are separated by an angular distance of $46\arcsec$, corresponding to 33,000 km separation between the lines of sight in the corona.  Most observations of natural radio sources have been performed at frequencies of $1-2$ GHz and, therefore, limited to heliocentric distances of $>5R_\odot$.  These observations demonstrate that, at higher frequencies, we can perform Faraday rotation measurements at shorter heliocentric distances.  Future observations at these frequencies may be able to perform sensitive measurements as close as $3-3.5R_\odot$.

\item We determined a mean rotation measure, $\overline{\rm{RM}}$, of $-0.69\pm0.33$ rad/m$^2$ and $4.83\pm0.31$ rad/m$^2$ for the northern and southern hotspots, respectively.  These rotation measures were lower than expected, but agree with the mean profiles of the rotation measure time series in Figure~\ref{fig:RM_Time_Series}.  The most striking feature of these mean profiles is their s-shape.  Although we could not correct for ionospheric Faraday rotation using CASA, we determined that the effects are $\lesssim1$ rad/m$^2$ and are therefore ignorable in all results presented here.

\item Model rotation measures were calculated from simple, analytical models of the coronal plasma and compared with the observations.  These models were similar to those employed in \cite{Mancuso&Spangler:2000} and \cite{Ingleby:2007} and consist of a radial magnetic field whose magnitude depends only on the heliocentric distance according to a single or dual power law. The model magnetic field reverses sign at a neutral line whose position is determined by potential field extrapolations of the photospheric field to a source surface located at $3.25R_\odot$ provided by the WSO.  The plasma density is represented by single or dual power laws in heliocentric distance.  Provision was also made for two-density models (Model 3, Section~\ref{Sec:Model_3}): a high density component corresponding to a streamer with a half-thickness of $12^{\circ}$ centered on the neutral line and a low density component representing the surrounding coronal hole.

The simplest analytic models, Models 1 and 2, successfully account for the sign of the rotation measure at the beginning of the observing session; however, they can not reproduce the trend of the rotation measure time series and they substantially overestimate the magnitude.  Model 3, which accounts for the streamer/coronal hole division and is most relevant to our observing conditions, produces improved agreement with observations.  It accounts for the sign and magnitude at the beginning of the observing session; however, it can not account for the observed trend in the rotation measure time series.  Previous work indicates that the simple neutral line geometry of Models 1 and 2 does a good job of reproducing the observed value of rotation measures at heliocentric distances of $5-10R_\odot$ and smaller $\beta_c$.  This work, however, suggests that at closer heliocentric distances and larger $\beta_c$, the finite thickness of the current sheet must be accounted for; it is no longer sufficient to consider the current sheet as an infinitely thin neutral line.

\item The extended, polarized structure of 3C228 permitted simultaneous measurements of the rotation measure along two lines of sight separated by $46\arcsec$ (33,000 km in the corona).  While most of the observations yielded no significant cases of rotation measure difference (termed differential Faraday rotation), three intervals with apparent non-zero differential Faraday rotation were observed: 15:37 - 16:05, 17:28 - 18:14, and 19:42 - 20:44 UTC.  These differential Faraday rotation measurements may be a result of coronal currents; using the method of \cite{Spangler:2007,Spangler:2009}, we determined the magnitude of these coronal currents should be on the order of $10^9-10^{10}$ A.  Assuming we did detect coronal currents, we reiterate the conclusion reported in \cite{Spangler:2007,Spangler:2009}: these values are several orders of magnitude below that which is necessary for significant coronal heating (assuming the Spitzer resistivity).

\item We examined the Faraday rotation measure fluctuations, $\delta\mathrm{RM}(t)$, present in the slowly varying time series by first removing the mean profiles (solid lines in Figure~\ref{fig:RM_Time_Series}).  No evidence was found for correlated fluctuations in the $\delta\mathrm{RM}(t)$ time series that could be interpreted as coronal Alfv{\' e}n waves.  Consequently, our data provide upper limits for the expected Faraday rotation fluctuations caused by magnetic field fluctuations.  At impact parameters of $\sim4.8R_\odot$, the $\delta\mathrm{RM}$ are no more than a few rad/m$^2$.  At observing frequencies near 2.3 GHz, the associated Faraday rotation fluctuations, $\delta\chi$, are therefore no more than a few degrees.  These upper limits are comparable to and, thus, not inconsistent with the theoretical models for Alfv{\' e}n wave heating of the corona by \cite{Hollweg:2010}.

\end{enumerate}


\acknowledgments
This work was supported at the University of Iowa by grants ATM09-56901 and AST09-07911 from the National Science Foundation. The space-based coronal occultation image is courtesy of the LASCO/{\it SOHO} consortium.  {\it SOHO} is a project of international cooperation between ESA and NASA.  We thank Kenneth Sowinski of the NRAO staff for helping us make measurements of the system temperatures for the VLA antennas.  We acknowledge and thank the referee of this paper, Michael Bird, for his thorough and authoritative review of this paper, which improved the presentation.  We also thank Kenneth Gayley for insightful comments that improved the paper.


\appendix

\section{Performance of the VLA at Solar Impact Parameters of $4.6 - 5.0 R_{\odot}$}\label{Sec:Tsys_discussion}
One of the goals of this project is to assess the performance of the VLA while pointing closer to the Sun (smaller ``impact parameters'') than in previous coronal Faraday rotation investigations.  One measure of array performance is the quality of the maps produced from the data.  This was discussed in Section~\ref{Sec:Imaging}.  However, a far more basic question is whether fundamental radio telescope parameters, particularly the system temperature $T_{sys}$, remain at values that are acceptable for sensitive polarimetry measurements when the antennas of the VLA are pointed $1\fdg0-1\fdg5$ from the Sun.  As mentioned in Section~\ref{sec:Observation}, this question motivated the test observations in \cite{Whiting&Spangler:2009}. The present project gave us an opportunity to check the results of \cite{Whiting&Spangler:2009} under the conditions of an actual coronal Faraday rotation observation.  

During the observations, we monitored and recorded the output of a synchronous detector that measured the ratio of a pulsed calibration diode to the system temperature.  The data were displayed on an operator console in the VLA control room.  We denote the reciprocal of this measurement as $SD$, and note that it is proportional to $T_{sys}$. Our primary observable was the ratio of $SD$ for an observation of 3C228 to that of $SD$ for a phase calibrator far from the Sun, shortly before or after the 3C228 observation. In the following discussion, we will refer to this ratio as the ``enhancement factor,'' $EF$.  The parameter $EF$ gives the ratio of the $T_{sys}$ including the effect of the Sun to that of the VLA antennas under normal operation.  Data were taken from two antennas and gave essentially the same results. 

Our values for $EF$ varied during the observing session, with an approximate range of 1.3 to 2.3.  Temporal variations during a session occur due to varying solar impact parameter, as well as the Sun moving through the distant sidelobes of the antenna as the source is tracked.  A mean value for $EF$ of about 1.7 was obtained from the data examined; on average, the system temperature for our observations was increased about 70\% above the nominal performance.  

The results presented in \cite{Whiting&Spangler:2009} predict $EF$ values of $ \simeq 1.30 - 1.45$. These estimates were read from Figure 3 of \cite{Whiting&Spangler:2009}.  The report of \cite{Whiting&Spangler:2009} therefore predicts somewhat smaller $T_{sys}$ increases than actually encountered.  

A portion of this difference (between predicted $30-45\%$ increases in $T_{sys}$ and the $\simeq$ 70\% encountered) can be explained by solar cycle effects, as anticipated in \cite{Whiting&Spangler:2009}.  The test observations of \cite{Whiting&Spangler:2009} were made on April 26, 2009, during the particularly pronounced solar minimum between solar cycles 23 and 24.  The observations discussed in the present paper were made on August 17, 2011, when solar activity indicators such as sunspot number had greatly increased.  

To determine the degree to which increased solar activity can account for these results, we checked values of the 2800 MHz solar flux density measured at the Penticton Radio Observatory.  We utilized data archived at the National Geophysical Data Center of the National Oceanographic and Atmospheric Administration (NOAA).  We used observed fluxes rather than those corrected to a standard distance of 1.00 a.u..  The 2800 MHz solar flux was 69.2 solar flux units (sfu)\footnote{One solar flux unit $= 10^{-22}$ W$-$m$^{-2}-$Hz$^{-1}$, or $10^4$ Jy.} on April 26, 2009, and 97.5 sfu on August 17, 2011.  These measurements therefore show a 41\% increase in the solar flux density between the two sets of observations.  

We assume a simple model in which  that $T_{sys}$ is the sum of the antenna temperature ($T_A$) and receiver temperature ($T_R$):
\begin{equation}
T_{sys} = T_{A} + T_{R}
\end{equation} 
In this case the enhancement factor is given by 
\begin{equation}
EF = \frac{T_{A}}{T_{R}} + 1
\end{equation} 
The \cite{Whiting&Spangler:2009} data would indicate that $T_A/T_R = 0.30 - 0.43$ at an impact parameter of $4.6 R_{\odot}$ in April, 2009.  Applying the factor of 1.41 to correct for the increased flux of the Sun at the (relatively) nearby frequency of 2800 MHz, and assuming the same value for $T_{R}$ in the two sessions leads to $T_A/T_R = 0.42 - 0.61$ for the August, 2011 observations, and corresponding enhancement factors $EF = 1.42 - 1.61$.  

These values are in the range of those measured, but do not seem to account for the high end of the range of encountered $T_{sys}$ values.  Our conclusion is that observed system temperatures that we observed at solar impact parameters of $4.6 - 5.0 R_{\odot}$ are roughly consistent with, but somewhat higher than those which would be indicated by the report of \cite{Whiting&Spangler:2009}.  The increased flux density of the Sun during this period, due to the change from solar minimum to the approach to solar maximum, accounts for part of this difference, although not all.

\newpage
\bibliographystyle{apj}
\bibliography{abbrev,Bibliography}

\end{document}